\documentclass{article}
\usepackage[utf8]{inputenc}

\usepackage[table,xcdraw]{xcolor}

\usepackage{tikz}
\usetikzlibrary{fit,positioning}
\usepackage[authoryear,round,colon]{natbib}
\usepackage{here}

\usepackage[linesnumbered,ruled,vlined]{algorithm2e}

\usepackage{amsmath, amssymb, amsthm}

\newtheorem{assumption}{Assumption}

\usepackage{pifont}
\newcommand{\cmark}{\ding{51}}%
\newcommand{\xmark}{\ding{55}}%

\edef\oldassumption{\the\numexpr\value{assumption}}

\usepackage{multirow}
\usepackage{hhline} % for table lines with colored cells
\usepackage{diagbox} % for diagonal line in table

\usepackage{subcaption}
\usepackage{makecell}

\usepackage{geometry}
\usepackage{listings} % R code 
\usepackage{fontawesome}

\usepackage{tabularx} % for tabularx macro

\newcommand\independent{\protect\mathpalette{\protect\independenT}{\perp}} 
\def\independenT#1#2{\mathrel{\rlap{$#1#2$}\mkern2mu{#1#2}}}
\usepackage{url}

\usepackage{multicol}

\usepackage[linesnumbered,ruled,vlined]{algorithm2e}

\usepackage{amsmath, amssymb, amsthm}

\newenvironment{assumptionbis}[1]
  {\renewcommand{\theassumption}{\ref{#1}b}%
   \addtocounter{assumption}{-1}%
   \begin{assumption}}
  {\end{assumption}}

\usepackage{multirow}
\usepackage{hhline} % for table lines with colored cells
\usepackage{diagbox} % for diagonal line in table

\usepackage{makecell}

\usepackage{here}

\usepackage{url}

\usepackage{dsfont} % for indicator function \mathds{1}

\usepackage{fontawesome}

\usepackage{tabularx}

\graphicspath{ {./figures/} }

\title{Generalizing treatment effects with incomplete covariates: identifying assumptions and multiple imputation algorithms}
\author{Imke Mayer\thanks{Institute of Public Health, Charit\'e -- Universit\"atsmedizin, Berlin, Germany (email: imke.mayer@inria.de). Corresponding author.}
  \and
    Julie Josse\thanks{PreMeDICaL, Inria Sophia-Antipolis, Montpellier, France (email: julie.josse@inria.fr).}
  \and
    Traumabase Group\thanks{Department of anesthesia and intensive care,  Beaujon hospital, AP--HP,  Clichy,  France.}
}
\date{\today}

\begin{document}

\maketitle

\begin{abstract}
We focus on the problem of generalizing a causal effect estimated on a randomized controlled trial (RCT) to a target population described by a set of covariates from observational data. Available methods such as inverse propensity sampling weighting are not designed to handle missing values, which are however common in both data sources. In addition to coupling the assumptions for causal effect identifiability and for the missing values mechanism and to defining appropriate estimation strategies, one difficulty is to consider the specific structure of the data with two sources and treatment and outcome only available in the RCT.
We propose three multiple imputation strategies to handle missing values when generalizing treatment effects, each handling the multi-source structure of the problem differently (separate imputation, joint imputation with fixed effect, joint imputation ignoring source information). As an alternative to multiple imputation, we also propose a direct estimation approach that treats incomplete covariates as semi-discrete variables. The multiple imputation strategies and the latter alternative rely on different sets of assumptions concerning the impact of missing values on identifiability. We discuss these assumptions  and assess the methods through an extensive simulation study.
This work is motivated by the analysis of a large registry of over 20,000 major trauma patients and an RCT studying the effect of tranexamic acid administration on mortality  in major trauma patients admitted to intensive care units (ICU). The analysis illustrates how the missing values handling can impact the conclusion about the effect generalized from the RCT to the target population.
 \vspace{12pt}
\vfill{}
\end{abstract}

\section{Introduction}

Observational and randomized trial data can provide different perspectives when evaluating an intervention or a medical treatment. Combining the information gathered from experimental and observational data is a promising avenue for medical research, because the knowledge that can be acquired from integrative analyses would not be possible from any single-source analysis alone. 
Such integrative analyses can be used to estimate the effect of a treatment on a specific target population different from the original trial population, to validate observational methods by emulating a trial \citep{hernan2016using}, to better estimate heterogeneous effects (which generally cannot be estimated from experimental data due to underpowered studies). 

Here, we are interested in the first case, where 
the results obtained from an experimental study on treatment effectiveness are only valid for the population studied in this experimental study, 
but do not reflect efficacy or effectiveness in a (slightly) different population or setting described by observational data that often better represent daily practice.
The experimental data or randomized controlled trial (RCT) is thus considered a biased sample of a target population and we wish to estimate the treatment effect on the target population represented by an observational study. Specifically, the effect is estimated using an RCT composed of covariates, treatment and outcome, whereas the observational study is composed only of covariates. 
There are a multitude of methods for generalizing a treatment effect, for detailed reviews we refer to \citet{Colnet2020} and \citet{degtiar2021review}. But all of these methods do not consider the problem of missing data which is ubiquitous in data analysis practice \citep{josse2018, Little2014}. 
We emphasize here that we focus on  covariates with sporadic missing values in both studies, while the outcome and treatment (available in the RCT) are assumed to be fully observed.

In certain cases, naive approaches such as complete-case analysis can yield unbiased treatment effect estimates \citep{bartlett_etal_2015}; however, in many settings, especially for observational data, the estimations are known to be biased since the complete-case observations are generally not a representative subsample of the population of interest \citep{Little2014}. Moreover, this approach is sometimes not even feasible since in high-dimensional settings the probability of having complete observations decreases rapidly \citep{zhu2019high}. There are a multitude of methods to handle missing values  \citep{Little2014, vanbuuren_2018, mayer_etal_2019}, such as maximum likelihood estimation or multiple imputations for parameter estimation and inference. These methods  make  assumptions on the mechanism that generated the missing values.
More recent works also consider the question of supervised learning with missing values which is a different problem from statistical inference of model parameters  \citep{LeMorvan2020}, but  supervised learning methods can be useful when estimating treatment effects \citep{chernozhukov_etal_2018, wager_athey_JASA2018}, e.g., to estimate non-parametric nuisance parameters on incomplete data for doubly robust estimators \citep{Mayer2020}. 
Missing data approaches for causal inference require coupling identifiability assumptions for the causal parameter with assumptions regarding the missingness mechanisms \citep{mattei_mealli_SMA2009, seaman_white_2014, yang_etal_2019,  kallus_etal_2018, blake_etal_2019, Mayer2020}. According to \citet{Mayer2020}, identifiability of the causal effect with missing values in observational studies can be ensured by 
adapting the causal inference assumptions to the missing values setting with an  \textit{unconfoundedness despite missing values} (UDM) assumption  \citep{Mayer2020,rosenbaum_rubin_JASA1984}. This assumption states that the unconfoundedness holds conditionally on the observed values and the missingness pattern. A formal definition of this assumption is provided in Section~\ref{sec:mia}.
However, these works only consider the case of a single dataset---or potentially multiple datasets with the same data distribution, i.e., sampled from the same population of interest---and do not treat the case of transporting or generalizing a treatment effect from an RCT to a target distribution defined through an observational dataset; the RCT representing a ``distorted'' population due to sampling or selection bias which arises due to eligibility requirements and also the trial setting. In practice, the observed distributions between the observational data and the RCT do not only differ due to the selection bias but may also differ in terms of missing values patterns.

To address the problem described above, our main contributions in this paper consist in: 

\begin{enumerate}
    \item defining several multiple imputation strategies adapted for integrating experimental and observational data (Section~\ref{sec:mi});
    \item proposing alternative identifiability assumptions which can be seen as an an extension of the \textit{unconfoundedness despite missingness} (UDM) assumption from the observational data case \citep{Mayer2020} and suggest adapted estimators (Section~\ref{sec:mia});
    \item assessing the performance of the proposed estimators and naive complete case estimators in an extensive simulation study %\footnote{The code for this simulation study as well as for the real-world data analysis is available as Supporting Information on the journal's web page (\url{http://onlinelibrary.wiley.com/doi/xxx/suppinfo})} 
    (Section~\ref{sec:simulations}), and can also be found in the following repository: \url{https://github.com/imkemayer/combined-incomplete-data};
    \item presenting the results of our analysis for the generalization of the effect of tranexamic acid on major trauma patients with traumatic brain injury (TBI) from an RCT to a target population described by observational data which has motivated this work (Section~\ref{sec:data-analysis}).
\end{enumerate}

\paragraph{Medical data analysis.}
Major trauma denotes injuries that endanger the life or the functional integrity of a person. The World Health Organization (WHO) has recently shown that major trauma including road-traffic accidents, interpersonal violence, falls, etc. remains a world-wide public health challenge and major source of mortality and handicap \citep{who2017}.
We focus on trauma patients suffering from a TBI. 
TBI is a sudden damage to the brain caused by a blow or jolt to the head and can lead to intracranial bleeding that can be observed on a computed tomography (CT) scan. 
Tranexamic acid (TXA) is an antifibrinolytic agent that limits excessive bleeding, commonly given to surgical patients.
Previous clinical trials showed that TXA decreases mortality in patients with traumatic \emph{extracranial} bleeding \citep{crash2}.
Such a result raises the possibility that it might also be effective in TBI, because \emph{intracranial} hemorrhage is common in TBI patients.
Therefore the question here is to assess the potential decrease of mortality in patients with intracranial bleeding when using TXA. 

While a multi-center international RCT with a majority of patients recruited in developing countries, ``CRASH-2", provides randomized data on this question, the ``Traumabase", an observational national registry is representative of the target population on which inference is desired. The latter contains information about critical care patients admitted in mainland France.
The details about these data are provided in Section~\ref{sec:data-analysis}.

\section{Background and notations}\label{sec:full-data}

We first introduce the notations, standard assumptions and estimators in the full data case, following the lines of \citet{nie2021covariate, Colnet2020}.

\subsection{Notations}

We consider that we have access to a trial sample (also called RCT) of $n>0$ independent and identically distributed (i.i.d.) observations drawn from  $\mathcal{P}_{trial}(X,Y(0),Y(1),A)$ where $X\in\mathcal{X}$ is a $p$-dimensional vector of covariates, $A$ denotes the binary treatment assignment, and  $Y(a)$ is the potential outcome for treatment level $a\in\{0,1\}$. 
We also have an observational sample of $m>0$ i.i.d observations drawn from $\mathcal{P}_{target}(X,Y(0),Y(1),A)$ but where only the covariates are observed.
The design considered is known as a \textit{non-nested design} and is illustrated in Figure~\ref{fig:typicalsituation-full}.

\begin{figure}[H]
\scriptsize
\setlength{\tabcolsep}{2.8pt}
\begin{center}
\begin{tabular}{ |c|c|ccc|c|c|c| } 
 \hline
\multirow{2}{*}{$i$}  &  \multirow{2}{*}{Set}   &\multicolumn{3}{c|}{Covariates} & Treatment & \makecell{Outcome \\under A=0} & \makecell{Outcome\\under A=1} \\
 &  &$X_1$ & $X_2$ & $X_3$ & $A$ & $Y(0)$ & $Y(1)$ \\ 
 \hline
1 & $Set_{\mathcal{R}}$ &1.1 & 20 & 5.4 & 1 &  23.4 & 24.1\\ 
$\dots$ & $Set_{\mathcal{R}}$  & & $\dots$ & & $\dots$ & $\dots$  & $\dots$ \\
$n-1$ & $Set_{\mathcal{R}}$  &-6 & 45 & 8.3 &  0 & 26.3 & 27.6 \\ 
$n$& $Set_{\mathcal{R}}$  &0 & 15 & 6.2 & 1 & 28.1 & 23.5 \\ 
$n+1$& $Set_{\mathcal{O}}$  & -2 & 52 & 7.1 & \texttt{NA} & \texttt{NA} & \texttt{NA} \\
$n+2$ & $Set_{\mathcal{O}}$  &-1 & 35 & 2.4 & \texttt{NA} &    \texttt{NA} & \texttt{NA}\\
$\dots$ & $Set_{\mathcal{O}}$ & & $\dots$ & & \texttt{NA} & \texttt{NA}  & \texttt{NA} \\
$n+m$& $Set_{\mathcal{O}}$  &-2 & 22 & 3.4 & \texttt{NA} &\texttt{NA} & \texttt{NA}\\
 \hline
\end{tabular}
% \hfill
\begin{tabular}{ |c|c|ccc|c|c| } 
 \hline
\multirow{2}{*}{$i$} & \multirow{2}{*}{Set}  &\multicolumn{3}{c|}{\makecell{Covariates\\}} & \makecell{Treatment\\} & \makecell{Outcome\\under A} \\
& & $X_1$ & $X_2$ & $X_3$ & $A$ & $Y$ \\ 
 \hline
1& $Set_{\mathcal{R}}$   &1.1 & 20 & 5.4 & 1 &  24.1 \\ 
$\dots$& $Set_{\mathcal{R}}$  & & $\dots$ & & $\dots$ & $\dots$   \\
$n-1$ &  $Set_{\mathcal{R}}$   &-6 & 45 & 8.3 & 0  & 26.3  \\ 
$n$&  $Set_{\mathcal{R}}$  &0 & 15 & 6.2 & 1 & 23.5  \\ 
$n+1$&  $Set_{\mathcal{O}}$  &-2 & 52 & 7.1 & \texttt{NA} & \texttt{NA}  \\
$n+2$ &  $Set_{\mathcal{O}}$  &-1 &  35& 2.4 & \texttt{NA} & \texttt{NA} \\
$\dots$& $Set_{\mathcal{O}}$  & & $\dots$ & & \texttt{NA} & \texttt{NA}   \\
$n+m$& $Set_{\mathcal{O}}$  &  -2 &  22 &3.4 & \texttt{NA} & \texttt{NA} \\
 \hline
\end{tabular}
\end{center}
\caption{Example of data structure in the full data problem setting with stacked trial and observational data. Left: complete but never observed underlying data. Right: observed data.}
\label{fig:typicalsituation-full}
\end{figure}

We emphasize that in this work, the observational data are assumed to be representative of the target population.
The density of $X$ in the target population (resp. trial population) is denoted $p_{target}(x)$ (resp. $p_{trial}(x)$). Throughout this work, we assume that there is a distributional shift, between the two populations which means that $\exists x \in \mathcal{X},\, p_{trial}(x) \neq p_{target}(x)$.  This makes it difficult to extrapolate average treatment effects from the trial to the target population.

We denote expectation with respect to the trial and the target distribution by $\mathbb{E}_{trial}$ and $\mathbb{E}_{target}$ respectively. and define the target population conditional average treatment effect (CATE):
\begin{equation}
\label{eq:cate}
\forall x\in\mathcal{X}\,,\quad \tau(x) = \mathbb{E}_{target} [Y(1) - Y(0) | X=x]\,,
\end{equation}

and the target population average treatment effect (ATE) 
$
\tau =\tau_{target}= \mathbb{E}_{target} [Y(1) - Y(0)],
$
while we define the RCT (or sample) average treatment effect as $\tau_{trial}=\mathbb{E}_{trial} [Y(1) - Y(0)]$.

We denote by $\mu_{a, target}(x)$ and $\mu_{a,trial}(x)$ the conditional response surfaces under treatment $a\in\{0,1\}$ in the general, i.e., the target, and in the RCT population, respectively:
$$
\mu_{a,target}(x) = \mathbb{E}_{target} [Y(a) | X=x]\,,\quad \mu_{a,trial}(x) = \mathbb{E}_{trial} [Y(a) | X=x]\,.
$$

and by $r(x)$ the density ratio that captures the assumed distribution shift between $\mathcal{P}_{trial}$ and $\mathcal{P}_{target}$:
$$
r(x) = \frac{p_{target}(x)}{p_{trial}(x)},\, \forall\, x\in\mathcal{X} \text{ such that } p_{trial}(x)\neq 0.
$$

\subsubsection{Alternative formulation of the generalization problem}\label{sec:alternative}

In the literature based on the potential outcomes framework, there are two equivalent formulations of the generalization problem. The distributional approach introduced above is used, for example, by \citet{nie2021covariate}. Alternatively, it is possible to adopt a sampling or selection process point of view by introducing a binary random variable $S\in\{0,1\}$ that indicates trial eligibility and willingness to participate\footnote{It does not correspond to a deterministic variable of whether or not the eligibility criteria are met, but rather illustrates that some individuals have more or less propensity to be chosen for the trial.}. This approach is taken, e.g., by \citet{Dahabreh2019} and \citet{Stuart2011}. The variable $S$ is only observed for individuals in the RCT while unknown for individuals from the target population sample. Indeed, for the individuals in the RCT, we know by design that $S=1$, but for the individuals from the target population, they could have been in the RCT according to their characteristics. Hence, both values of $S$ can occur in the observational sample. It is generally assumed in this context that if an individual meets the eligibility criteria and in principle is willing to participate, then the actual trial participation is random and does not depend on other factors, observed or unobserved. 
Trial eligibility and willingness to participate are assumed to depend to some extend on the covariates $X$, so that there is a shift in the distributions of the trial and the target populations.
The following assumptions for identifiability of the ATE on the target population can also be formulated using this alternative notation $S$ (see, e.g., \citet{Colnet2020}).

\subsection{Assumptions for identifiability of the ATE on the target population in the full data case}\label{sec:idenassump}

The main identifiability assumptions that allow for generalizing an ATE from the RCT onto a target population are as follows:

\subsubsection{Internal validity of the RCT}

\begin{assumption}[Consistency of potential outcomes]\label{a:consist} $Y=A\,Y(1)+(1-A)\,Y(0).$
\end{assumption}

\begin{assumption}[Treatment randomization]\label{a:random}
$\mathbb{E}_{trial}[A|Y(a), X=x] = \mathbb{E}_{trial}[A], \text{ for all } x\in\mathcal{X} \text{ and } a=0,1.$
\end{assumption}
Moreover, for simplicity of the following presentation, we assume a constant treatment assignment propensity in the RCT, i.e., $e_{trial}(x)=Pr(A = a \, |\, X = x)=0.5$ for all $x$. %such that $p_{trial}(x)>0$

\subsubsection{Generalizability of the RCT to the target population}

\begin{assumption}[Transportability of the CATE\footnote{In the literature, this assumption is sometimes also referred to as \textit{exchangeability with respect to trial eligibility and willingness \citep{Stuart2011, degtiar2021review}}.}]\label{a:trans-2}
$\mathbb{E}_{target}[Y(1)- Y(0)|X=x]= \mathbb{E}_{trial}[Y(1)- Y(0)|X=x]$ for all $x\in\mathcal{X}$.
\end{assumption}

\begin{assumption}[Support inclusion]\label{a:pos}        $supp(\mathcal{P}_{target}(X)) \subseteq supp(\mathcal{P}_{trial}(X)),$ where $supp(\mathcal{P})$ denotes the support of distribution $\mathcal{P}$.
\end{assumption}

Assumption \ref{a:trans-2} states that the conditional treatment effect is stable across populations, therefore it implies that $X$ contains all treatment effect modifiers (in this case with respect to the absolute value of the difference in potential outcomes, i.e. on the absolute scale) that differ in distribution between the trial and target populations. 
With {Assumption \ref{a:pos}} we formalize an implication of the non-nested design, namely that the support of the observational sample is included within the support of the RCT. Without this support inclusion assumption, we cannot hope to generalize the effect to units from the target population which have a zero probability of being eligible and of participating in the trial. For example, if we assume that age is a treatment effect modifier and that the target population is composed mostly of older individuals than in the trial, the support inclusion assumption requires that older individuals had a non-zero chance of being eligible and included in the trial \citep{Stuart2011}.

\subsection{Estimators in the full data case}\label{sec:full-data-estimators}

The covariate distribution of the RCT population $\mathcal{P}_{trial}$ is generally different from that of the target population $\mathcal{P}_{target}$; therefore, $\tau_{trial}$ is different
from $\tau$, and an estimator based solely on the RCT, such as the difference in means estimator $\widehat \tau_{trial} = \frac{1}{n_1}\sum_{i=1}^{n} A_iY_i - \frac{1}{n_0}\sum_{i=1}^{n} (1-A_i)Y_i$, where $n_a=\sum_{i=1}^n \mathbf{1}_{\{A_i=a\}}$, is biased
for the ATE of interest $\tau$. 
Under the previous identifiability assumptions, different estimators are available to estimate the ATE $\tau$ (identification formula as well as their derivation are provided in Appendix~\ref{appendix:full-data}): the inverse probability of sampling weighting (IPSW) proposes to reweight, using the density ratio $r(x)$, the RCT sample so that it ``resembles'' the target population with respect to the observed ``shifted" covariates.
The conditional outcome-based estimator proposes to model the conditional outcomes with and without treatment in the RCT, and then to apply the model to the target population of interest. Doubly robust approaches leverage combinations of the former two, improving the robustness and efficacy. 

\paragraph{Inverse probability of sampling weighting (IPSW).}

This estimator is defined as the weighted difference of average outcomes between the treated and control groups in the trial \citep[e.g.,][]{Cole2010, Stuart2011}. The observations are weighted by the density ratio $r(x)$ to account for the shift of the covariate distribution from
the RCT sample to the target population. 
The IPSW estimator can be written as follows: 
\begin{equation}
\label{eq:ipsw}
\widehat{\tau}^{IPSW}= \frac{2}{n} \sum_{i=1}^{n} \widehat{r}(X_i) Y_i\left( 2A_i - 1 \right)\,,
\end{equation}
where $\widehat r$ is an estimate of $r$. The IPSW estimator is consistent as long as $r$ is consistently estimated by $\widehat r$. Note that the density ratio can be estimated using different strategies, e.g., by moment matching \citep{nie2021covariate} or, assuming that $\widehat r(x) = \frac{n}{m}\frac{\widehat{Pr}(i\in Set_{\mathcal{O}}| X_i=x)}{\widehat{Pr}(i\in Set_{\mathcal{R}}| X_i=x)}$ \citep{colnet_etal_2021}, the density ratio can be estimated from the data using, e.g., a logistic regression model for the indicator $\mathds{1}_{Set_{\mathcal{R}}}$ on covariates $X$.

\paragraph{Conditional outcome-based estimation.} \label{sec:cond-outcome}

This approach fits regressions of the conditional response surfaces among trial participants. 
Marginalizing these regressions over the covariate distribution of the observational data, 
gives the corresponding expected outcome \citep[e.g.,][]{Dahabreh2019c}.
This outcome-model-based estimator, an instantiation of the so called g-formula estimator \citep{Robins1986} to the case of treatment effect generalization, is then defined as: 
\begin{equation}
\label{eq:g-formula}
\widehat \tau^{CO} = \frac{1}{m}\sum_{i=n+1}^{n+m}\left(\widehat \mu_{1, trial}(X_i) - \widehat \mu_{0,trial}(X_i)\right),
\end{equation}
where $\widehat \mu_{a, trial}(X_i)$ is an estimator of $\mu_{a, trial}(X_i)$.
If the model is correctly specified, then the estimator is consistent.

\paragraph{Doubly robust estimators.} \label{sec:DR}

The density ratio and outcome regression models used in the first two estimators can be combined to form
an augmented IPSW estimator (AIPSW):
\begin{equation}
\begin{split}
\widehat{\tau}^{\mathrm{AIPSW}}= \frac{2}{n} \sum_{i=1}^{n}  \widehat r(X_i)\left[A_{i}\left\{ Y_{i}-\widehat\mu_{1, 1}(X_{i})\right\} -(1-A_{i})\left\{ Y_{i}-\widehat\mu_{0, 1}(X_{i})\right\} \right] \\
+\frac{1}{m}\sum_{i=n+1}^{m+n}\left\{ \widehat\mu_{1, trial}(X_{i})-\widehat\mu_{0, trial}(X_{i})\right\} . 
\end{split}
\label{eq:aipsw}
\end{equation}
It is doubly robust, i.e., consistent and
asymptotically normal when either one of the two estimators $\widehat r(X)$ and $\widehat\mu_{a,trial}(X)$ $(a=0,1)$ is consistent \citep{Dahabreh2019}.

\paragraph{Calibration weighting.}\label{sec:CWmethod}

IPSW is likely to be unstable if some of the estimated weights are very small \citep{Dahabreh2019}. To resolve the instability of IPSW calibration weighting \citep{dong2020integrative} has been proposed. 
They calibrate, i.e., they balance 
the covariates between the RCT sample and the target population: usually the balance is enforced on the first and second moments of the covariates such that the weighted mean and variance for each variable in the RCT match the ones in the observational data. 
More precisely, in order to calibrate, they assign an
entropy-balancing weight $\omega_{i}$ to each subject $i$ in the RCT sample obtained by solving an optimization problem:
\begin{align}
\underset{\omega_1,\ldots,\omega_n}{\text{min}} & \sum_{i=1}^{n}\omega_{i}\log \omega_{i},\label{eq:optQ}\\
\text{subject to}\quad & \omega_{i}\ge0,\;\text{for all}\;i,\nonumber
\sum_{i=1}^{n}\omega_{i}=1, \text{ and } \sum_{i=1}^{n} \omega_{i}\,\mathbf{g}(X_{i})=\widetilde{\mathbf{g}},\;\text{(balancing constraint) \nonumber}\label{eq:calibration constraints}
\end{align}
where $\widetilde{\mathbf{g}}=m^{-1}\sum_{i=n+1}^{m+n}\mathbf{g}(X_{i})$
is a consistent estimator of $\mathbb{E}_{target}[\mathbf{g}(X)]$ from the observational sample. 
The balancing constraint calibrates the covariate distribution of the RCT sample to the target population in terms of $\mathbf{g}(X)$. The objective function in \eqref{eq:optQ} is the negative entropy of the calibration weights; thus, minimizing this criterion ensures that the empirical distribution of calibration weights are not too far away from the uniform, such that it minimizes the variability due to heterogeneous/extreme weights. 
Based on the calibration weights, the CW estimator is then defined as 
\begin{equation}
\widehat{\tau}^{\mathrm{CW}}=2\sum_{i=1}^{n}\widehat{\omega}_{i} Y_i (2A_i-1).\label{eq:cbss0}
\end{equation}
This estimator is doubly robust in that
it is a consistent estimator for $\tau$ if either the obtained weights $\widehat{\omega}_{i}$ are proportional to the density ratio $r(X)$ \citep{chu2022targeted}\footnote{This is the case for example if $Pr(S=1|X)$ follows a logistic regression model, where $S$ corresponds to the random variable from the selection process approach briefly mentioned in Section~\ref{sec:alternative}.}, 
or if the CATE \eqref{eq:cate} is linear in the calibration constraint.

\subsection{Missing values mechanisms}\label{sec:missing-values}

In the taxonomy proposed by \citet{rubin_1976}, missing data mechanisms can be \textit{missing completely at random} (MCAR) when the missingness is independent of the data, \textit{missing at random} (MAR) when the missingness depends only on observed values, or \textit{missing not at random} (MNAR) when missingness can depend on unobserved values. The first two mechanisms are ``ignorable'' in that, the missing data mechanism is separable in the full data likelihood function, so missingness can be ``ignored''. 
More formally, we denote the response pattern of the $i$-th sample as  $R_i\in\{0,1\}^p$ such that $R_{ij}=1$ if $X_{ij}$ is observed and $R_{ij}=0$ otherwise. For all response patterns $r$ and $X=(X_{obs(r)}, X_{mis(r)})$ the partition of the data in realized observed and missing values given a specific realization of the pattern, we define the missing values mechanisms as follows.
\begin{align}
\label{eq:mcar}
\text{(MCAR in the trial)} & \quad \forall\, r\in\{0,1\}^p,\, Pr(R=r|X, A, Y) = Pr(R=r) \\
\label{eq:mcar-target}
\text{(MCAR in the target)} & \quad \forall\, r\in\{0,1\}^p,\, Pr(R=r|X) = Pr(R=r) \\
\label{eq:mar}
\text{(MAR in the trial)} & \quad \forall\, r\in\{0,1\}^p,\, Pr(R=r|X, A, Y) = Pr(R=r|X_{obs(r)}, A, Y)\\
\label{eq:mar-target}
\text{(MAR in the target)} & \quad \forall\, r\in\{0,1\}^p,\, Pr(R=r|X) = Pr(R=r|X_{obs(r)})
\end{align}

For more details and nuanced presentation of the missing values mechanisms and related assumptions we refer to \citet{mealli2015clarifying}.
In Appendix~\ref{appendix:missing-data}, we give an illustration in Figure~\ref{fig:typicalsituation-na} of the considered incomplete data, similarly to the example of the complete data case of Figure~\ref{fig:typicalsituation-full}.

\section{Multiple imputation under standard identifiability assumptions}\label{sec:mi}

Under the standard identifiability assumptions, namely Assumptions \ref{a:consist}-\ref{a:pos}, and separate assumptions on the missingness mechanism as described in the previous section, multiple imputation can be an adequate strategy to address the challenge of generalizing the average treatment effect with incomplete covariates.

\subsection{General concept}

Multiple imputation (MI) is one of the most powerful approaches to estimate parameters and their variance from incomplete data \citep{Little2014, Kim2013b, Schafer2010}. For a single dataset, it consists in generating $M$ plausible values for each missing entry, which leads to $M$ completed datasets. Then, an analysis is performed on each imputed data set $m=1,\,\dots,\,M$, to get an estimate for the parameter of interest, say $\theta$ as $\widehat \theta^{m}$ and an estimate of its variance  $\widehat V^m(\widehat \theta^m)$ and the results are combined using \citet{Rubin1987}'s rules to get correct inference with missing values, namely confidence intervals with the appropriate coverage.

\subsection{Adapted multiple imputation for multiple data sources with different data design}

For our problem, there are multiple possibilities to derive a multiple imputation strategy to generalize a treatment effect. This is due to the multi-source structure of the data and the fact that there is an additional complication due to the number of variables are not the same in the RCT and in the observational study. Indeed, we assume that the observational study does not include treatment and outcome but only covariates. We stress again that we assume missing values only occur in the covariates of both data. 
We suggest and describe three strategies to tackle this problem:

\begin{enumerate}
    \item Within-study multiple imputation:
    \begin{enumerate}
        \item \sloppy Multiple imputation of the RCT: Impute $M$ times the covariates of the RCT using $(X_i, A_i, Y_i)_{i:\, i\in Set_{\mathcal{R}}}$.
        \item Multiple imputation of the observational data: Impute $M$ times the covariates of the observational data using only the covariates  $(X_i)_{i:\,i\in Set_{\mathcal{O}}}$.
        \item Create $M\times M$ complete tables by concatenating all possible combinations of imputed RCT and observational data. Estimate the treatment effect on every combination using any complete case estimator as in Section \ref{sec:full-data} and aggregate these estimations using Rubin's rules.  
    \end{enumerate}
    \item   Ad-hoc joint covariates multiple imputation, ignoring the information on the data sources, i.e., ignoring the source indicator $\mathds{1}_{ Set_{\mathcal{R}}}$:
    \begin{enumerate}
        \item Impute $M$ times the joint datasets (the concatenation of the covariates from the RCT and the ones from the observational study) with covariates $X$.
         \item Concatenate the outcome $Y$ and treatment $A$ for each imputed RCT individual. 
        \item Compute the $M$ treatment effect estimators using any complete case estimator as in Section \ref{sec:full-data} and aggregate them using Rubin's rules.
    \end{enumerate}
    \item 
    Joint covariates multiple imputation, modeling the group variable as a fixed effect, i.e., explicitly use the indicator $\mathds{1}_{Set_{\mathcal{R}}}$ for the corresponding ``group''/``source'' during the imputation:
    \begin{enumerate}
        \item Impute $M$ times the joint datasets with covariates $X$ and model the source indicator $\mathds{1}_{Set_{\mathcal{R}}}$. 
         \item Concatenate the outcome $Y$ and treatment $A$ for each imputed RCT individual. 
        \item Compute the $M$ treatment effect estimators using any complete case estimator as in Section \ref{sec:full-data} and aggregate them using Rubin's rules.
    \end{enumerate}
\end{enumerate}
A schematic illustration of these three strategies is given in Figure~\ref{fig:mi} in the Appendix.
The first strategy has the advantage that it takes into account the outcome $Y$ and treatment $A$ which are dependent variables of the covariates $X$,  when imputing the covariates of the RCT as suggested  by  \citet{leyrat_etal_2019, seaman_white_2014, mattei_mealli_SMA2009} as it models the entire joint distribution. 
The other strategies only consider the covariates $X$. The second strategy solely relies on the relationships between the covariates $X$ and the assumption that these are stable across the data sources, i.e., $Cov_{target}(X)= Cov_{trial}(X)$, where $Cov(Z)$ denotes the covariance matrix of the random vector $Z$. Strategy 3 can be seen as a fixed effect method, where the variable $Q$ is included as a variable in the imputation model which allows, e.g., in case of multiple imputation with conditional regression  models, to impute according to an analysis of covariance model (to take into account differences between the means of the variables within each data source).

All three strategies can be implemented easily using the R package mice \citep{vanbuuren_2018} which uses conditional models such as linear and logistic regressions (by default) to perform multiple imputation.

Multiple imputation is especially suited if the missing values are ignorable as described in Section~\ref{sec:missing-values} and if the identifiability assumptions of the ATE $\tau$ in the full data case are met, namely Assumptions \ref{a:trans-2} and \ref{a:pos}. In case of nonignorable missing values, multiple imputation is still possible but requires knowledge about the generating process for the missing values to explicitly model the response or missingness pattern \citep{vanbuuren_2018}.
There is no clear rule about the number of multiple imputations to achieve good performances, however an accepted rule of thumb is to choose the number of imputations to be similar to the percentage of incomplete cases \citep{Hippel2009} or to the average percentage of missing data \citep{vanbuuren_2018}. In the considered settings and for the deployed imputation strategies, increasing the number of imputations from 10 to 50 does not improve the final result.

\section{Missing incorporated in attributes under alternative identifiability assumptions}\label{sec:mia}

An alternative to handle missing covariates values consists in modifying the identifiability assumptions so that they directly handle missing values but do not necessarily require assumptions on the missing values mechanism. 
This can be seen as an advantage as it possibly allows for MNAR data, but the new identifiability assumptions may be more difficult to satisfy than in the full data case. More precisely, to enable generalizing the RCT to the target population, we extend the work of \citet{Mayer2020}
 who adapt the unconfoundedness assumption to the incomplete covariates in order to identify the (average) treatment effect in the observational data case. 
First, we introduce an additional notation, required for the following approach:
We denote the matrix of observed covariates with $X_i^{*} \triangleq X_i \odot R_i  + \texttt{NA} \odot (\mathbf 1- R_i)$, with $\odot$ the element-wise multiplication and $\mathbf 1$ the matrix filled with 1,  so that $X_i^{*}$ takes its value in the half discrete space
${\mathcal X^{*}} \triangleq \underset{1\leq j\leq |\mathcal{X}|}{\times}\{\mathcal{X}_j\cup \{\texttt{NA}\}\}$. 
Now we can replace the previous Assumptions \ref{a:trans-2} and \ref{a:pos} with the following alternatives.

\begin{assumptionbis}{a:trans-2}[Transportability of the CATE, conditionally independent selection (CIS)]\label{a:trans-2-na}    
\sloppy $\mathbb{E}_{target}[Y(1)- Y(0)|X^*=x^*]= \mathbb{E}_{trial}[Y(1)- Y(0)|X^*=x^*]$ for all $x^*\in\mathcal{X}^*$.
\end{assumptionbis}

\begin{assumptionbis}{a:pos}[Support inclusion]\label{a:pos-na}  
 $supp(\mathcal{P}_{target}(X^*)) \subseteq supp(\mathcal{P}_{trial}(X^*)).$
\end{assumptionbis}

The intuition behind these alternative assumptions is to assume that instead of requiring transportability of the CATE conditionally on all covariates, we only require transportability of the CATE conditionally on the \textit{observed} information, meaning the observed values and the pattern of missing values. To further elucidate this assumption, we temporarily take the alternative selection point of view introduced in Section~\ref{sec:alternative}: The CIS assumption, which can also be written as $S\independent Y(1), Y(0) | X$, may be plausible in a context where trial eligibility (and willingness to participate) are defined via a set of sufficient conditions that are not all necessary conditions. In practice, one could imagine a ``check list'' of $L$ conditions and that it is necessary to fulfill at least $l< L$ of these to qualify for the trial. More specifically, if we consider five covariates $X_1,\dots, X_5$ and assume that there exist different possibilities to be included in the trial, e.g., a condition on $X_1, X_2$ and $X_3$ regardless of $X_4$ and $X_5$ and another alternative condition on $X_2$ and $X_5$. In the case of such a design, the CIS assumption is potentially a suited assumption to generalize a treatment effect onto another population.
Taking the example given in Figure~\ref{fig:typicalsituation-na} (in Appendix~\ref{appendix:missing-data}), for observation $1$, only $X_1$ and $X_2$ and the fact that $X_3$ is unobserved qualify individual $1$ for the trial and willingness to participate, while for observation $2$, only $X_1$ and $X_3$ and the fact that $X_2$ is missing contribute to the value taken by $S$, etc.

Similar to the UDM assumption from \citet{Mayer2020} which states that $\{Y(0), Y(1)\} \independent A \,|\, X^*$ in the observational data case, i.e., the treatment assignment is independent of the treatment effect conditionally on the observed values and the missingness pattern, Assumption \ref{a:trans-2-na} can be replaced by two sufficient assumptions, using the selection indicator $S$ introduced in Section~\ref{sec:alternative}: Assumption \ref{a:trans-2} and $S \independent X \,|\, X^*$, thus the term \textit{conditionally independent selection}. 

\subsection{Generalized estimators}\label{sec:mia-estimators}

Similar to \citet{rosenbaum_rubin_JASA1984, dagostino_rubin_JASA2000, Mayer2020}, we define the generalized conditional response surfaces $\mu_{a,target}^{*}$ and $\mu_{a,trial}^{*}$ as follows:

\begin{equation}
\label{eq:generalized-outcome}
    \mu_{a,target}^{*}(x^*)  = \mathbb{E}_{target}[Y(a) | X^{*} = x^{*}], \quad
    \mu_{a,trial}^{*}(x^*)  = \mathbb{E}_{trial}[Y(a) | X^{*} = x^{*}].
\end{equation}

The resulting estimators are then formed analogously to the estimators in the full data case, by substituting the corresponding nuisance parameters with their generalized counterparts. More explicitly, the outcome-model-based estimator defined by \eqref{eq:g-formula} becomes 
\begin{equation}
\label{eq:g-formula_star}
\widehat \tau^{CO,*} = \frac{1}{m}\sum_{i=n+1}^{n+m}\left( \widehat\mu^*_{1, trial}(X_i^*) - \widehat\mu^*_{0,trial}(X_i^*)\right).
\end{equation}

Fitting the new nuisance  \eqref{eq:generalized-outcome} is not straightforward, since these require to fit a separate regression model for each possible pattern $r$ of missing values \citep{LeMorvan2021}. For example, if we have three incomplete covariates $X_1, X_2, X_3$, with the $2^3$ patterns of missing values, this means we need to fit a separate regression on $\{X_1, X_2, X_3\}$, on $\{X_1, X_2\}$, on $\{X_2, X_3\}$, on  $\{X_1\}$, etc. We can see from this example that this is not possible in moderate and high dimensions with classical regression methods. 

\paragraph{Nonparametric estimation.}
We propose to estimate the generalized nuisance components via 
random forests \citep{breiman2001random,athey_etal_AS2019}, with missing data handled
using the \textit{missing incorporated in attributes} (MIA) method of \citet{twala_etal_2008}. Indeed, as noted already by \citet{athey_etal_AS2019, Mayer2020}, many modern machine learning methods, including tree ensembles and neural networks, can  be adapted to this context and thus readily handle
missing data and enable direct fitting of the generalized regression models \eqref{eq:generalized-outcome}, for detailed consistency results we refer to \citet{LeMorvan2021}. The resulting IPSW, CO and AIPSW estimators will be denoted by \smash{$\hat{\tau}_{MIA}^{IPSW,*}$}, \smash{$\hat{\tau}_{MIA}^{CO,*}$}, and \smash{$\hat{\tau}_{MIA}^{AIPSW,*}$} respectively.

In random trees, the MIA approach extends the classical splitting rules such that missing
values are incorporated in the splitting criterion.
More specifically, consider splitting on the $j$-th attribute
and assume that for some individuals, the value of $X_j$ is missing, MIA treats the missing values as
a separate category or code and considers the following splits:
\begin{enumerate}
\item $\{i: X_{ij} \leq t \text{ or } X_{ij} \text{ is missing}\}$ vs. $\{i:X_{ij} > t\}$
\item $\{i: X_{ij} \leq t\}$ vs. $\{i:X_{ij} > t \text{ or } X_{ij} \text{ is missing}\}$
\item $\{X_{ij}  \text{ is missing}\}$ vs. $\{X_{ij} \text{ is observed}\}$,
\end{enumerate}
for some threshold $t$.
The MIA approach does not seek to model why some features are unobserved; instead, it
simply tries to use information about missingness to make the best possible splits for modeling
the desired outcome. Thus the MIA strategy works with arbitrary missingness mechanisms
and does not require the missing data to follow a specific mechanism. This MIA approach for (generalized) random forests is implemented in the R package \texttt{grf} \citep{grf} which is also used in the simulation part of this work presented in Section~\ref{sec:simulations}.

\paragraph{Parametric alternative.}
Parametric estimation is however possible in the case of logistic and linear regression models. This is based on work by \citet{jiang_etal_2018} and \citet{Schafer2010} for logistic and linear regressions with missing covariates. 
The regression functions $\mu_{a,trial}^{*}$ defined in \eqref{eq:generalized-outcome} that take in incomplete covariates $x^*$ are estimated via EM \citep{dempster_etal_1977}.
The resulting IPSW, CO and AIPSW estimators will be denoted by \smash{$\hat{\tau}_{EM}^{IPSW,*}$}, \smash{$\hat{\tau}_{EM}^{CO,*}$}, and \smash{$\hat{\tau}_{EM}^{AIPSW,*}$} respectively.
The details of this approach are given in the Appendix~\ref{appendix:methods}. 

However, a major limitation of this approach is that, in addition to the alternative identifiability assumptions \ref{a:trans-2-na} and \ref{a:pos-na}, in order to justify the use of the EM algorithm, one typically needs to make further assumptions
on the missing value mechanism; in particular, this approach assumes the MAR mechanism, i.e., \eqref{eq:mar} and \eqref{eq:mar-target}.
In other words, although we did not require the missing at random assumption to identify $\tau$, this assumption is
used for consistent parametric estimation of the generalized conditional regression models $\mu^*_{a,trial}$.

\section{Simulations}\label{sec:simulations}

We conduct a detailed simulation study to assess the performance of the previously introduced estimators to handle missing values. This controlled study allows to quantify the impact of different missing values mechanisms and identifiability assumptions on the final estimate for the ATE $\tau$. The code for this simulation study as well as for the medical data analysis (Section~\ref{sec:data-analysis}) has been written and run using R version 4.0.2 (platform: x86\_64-apple-darwin17.0, 64-bit) with packages mice\_3.14.0, misaem\_1.0.1, genRCT\_0.1.0, grf\_2.0.2, caret\_6.0-90.\footnote{Computation of all estimators for a single replication with fixed sample size ($(n,m)=(1000,10000)$), fixed missingness mechanism and proportion and fixed identifiability assumption, computing all estimators takes approximately 20 minutes.}

\subsection{Data generation}

\subsubsection{Standard assumptions for causal identifiability  and  missing values mechanisms}

In this setting, we consider that Assumptions \ref{a:consist}-\ref{a:pos} hold. If additionally, missing values are generated using a MAR mechanism (\eqref{eq:mar} and \eqref{eq:mar-target}), we expect the multiple imputation based methods proposed in Section \ref{sec:mi} to perform well. The data simulation is implemented in two steps, using the selection indicator $S$ from the selection process approach mentioned in Section~\ref{sec:alternative} which provides a more transparent way to simulate adequate data: (1) a large covariate sample is drawn from the target population, then a trial sample is generated out of this sample using a model for trial eligibility and participation (selecting observations with $S=1$ and discarding all other observations); (2) the target population sample is drawn from the target population independently from the first sample. 
This two-step simulation approach results in a covariate shift between the trial and target population as assumed throughout this work. We consider the selection model generated either as a logistic model as follows
\begin{equation}
\label{eq:selection-model}
\operatorname{logit}\left\{Pr(S=1|X)\right\}=-3.1-0.5 X_{1}-0.3 X_{2}-0.5 X_{3}-0.4 X_{4},
\end{equation}
or according to the following non-linear model:
\begin{align}
\begin{split}
\label{eq:selection-model-nonlin}
\operatorname{logit}\left\{Pr(S=1|X)\right\}=&-2.95-0.5 |X_{1}|\sin(X_{1})-0.3 |X_{2}|\overline{X_{2}} -0.75X_{3} \\
& - 0.5 X_{3}|X_{1}|\sin(X_{1})-0.4 |X_{4}|\overline{X_{4}},
\end{split}
\end{align}
where every $X$ is drawn from a multivariate normal distribution with mean $1$ and covariance matrix $\Sigma$ such that $\Sigma_{ij}=\left\{\begin{array}{ll} 1& \text{ if } i=j\\ 0.6 &\text { if } i \neq j \end{array}\right.$ to have correlated covariates. 
The outcome is generated according to either the linear model below such that $X_1$ is a treatment effect modifier and the true ATE $\tau$ is set to 27.4,

\begin{equation}
\label{eq:outcome-model}
Y(a)=-100+27.4 a X_{1}+13.7 X_{2}+13.7 X_{3}+13.7 X_{4}+\epsilon \quad \text{ with }\epsilon \sim \mathcal{N}(0,1),
\end{equation}
or to the following non-linear model with the true ATE set to 58.9.

\begin{align}
\label{eq:outcome-model-nonlin}
\begin{split}
Y(a)=& -100+27.4 a (|X_{1}|\sin(X_{1})+1.5) +13.7 |X_{2}|\overline{X_{2}} \\
& +20.55 X_{3} +13.7 X_{3}|X_{1}|\sin(X_{1}) +13.7 |X_{4}|\overline{X_{4}}+\epsilon \quad \text{ with }\epsilon \sim \mathcal{N}(0,1).
\end{split}
\end{align}

Missing values in the covariates are generated as follows, defining different models for the response indicator $R$:

\begin{enumerate}
\item Missing values can occur in all four covariates.
\item Proportion of missing values in each incomplete covariate: 20\%.
\item The missing values mechanism can be MCAR or MAR and is implemented in the \texttt{produce\_NA} function\footnote{\url{https://rmisstastic.netlify.app/how-to/generate/missSimul.pdf}} proposed by \citet{mayer_etal_2019}. Details on the different mechanisms can be found in Appendix~\ref{appendix:results}.
\end{enumerate}

The simulation design is summarized by Algorithm~\ref{algo:simulation-design-standard} provided in Appendix~\ref{appendix:results}.

\subsubsection{Alternative identifiability assumptions for treatment effect generalization}

We also generate data according where the CIS assumption \ref{a:trans-2-na} is met. In such scenarios, we expect that the methods described in Section \ref{sec:mia} will work best.
The main difference with the previous setting of simulation, lies in the definition of the selection model, the outcome model remaining unchanged. In this setting we also follow a two-step simulation scheme. 

In order to simulate data under the CIS assumption \ref{a:trans-2-na}, we need to modify the definition(s) of the selection model such that it becomes pattern-dependent.\footnote{As a concrete example, in the critical care context, missing values of the generally easily measurable pre-hospital blood pressure and heart frequency are generally indicators of severity of a trauma patient as patients with cardiac arrest or severe external hemorrhage require immediate care on the site of the accident that makes the measurements of these variables difficult or even impossible. Therefore, if pre-hospital assessed severity were a trial eligibility criterion, the missingness patterns of these variables would be relevant for determining $S$. But for other patients the measured values of these variables could also be of relevance to decide upon trial eligibility.}
\begin{equation}
\label{eq:selection-model-cis}
\operatorname{logit}\left\{Pr(S=1|X)\right\}=-2.5-0.5 X_{1}\odot R_{1}-0.3 X_{2}\odot R_{2}-0.5 X_{3}\odot R_{3}-0.4 X_{4}\odot R_{4},
\end{equation}
\begin{align}
\label{eq:selection-model-cis-nonlin}
\begin{split}
\operatorname{logit}\left\{Pr(S=1|X)\right\}=&-2.1-0.5 (|X_{1}|\sin(X_{1})+1.5)\odot R_{1}-0.3 |X_{2}|\overline{X_{2}}\odot R_{2}\\
& -0.75X_{3}\odot R_{3} - 0.5 X_{3}|X_{1}|\sin(X_{1})\odot R_{3}-0.4 |X_{4}|\overline{X_{4}}\odot R_{4},
\end{split}
\end{align}
In Appendix~\ref{appendix:results}, we summarize the simulation design and the CIS assumption in Algorithm \ref{algo:simulation-design-cis}.

\subsection{Estimation methods}

We consider different scenarios of data generating processes by varying the type of missing values (MCAR, MAR, MNAR), CATE transportability assumption (standard or CIS), as well as the number of observations.
We compare the following methods to handle missing values (the following acronyms are identical to the method labels used in  Figures \ref{fig:standard-bias-nsim100} --\ref{fig:cis-bias-nsim100}:
\begin{enumerate}
\item Full data: we apply the standard full data estimators from Section~\ref{sec:full-data} on the full data before introducing missing values (this would serve as a reference).
\item Complete cases (CC): we apply the standard full data estimators from Section~\ref{sec:full-data} on the complete observations extracted from the incomplete data (by deleting observations with missing values).\footnote{Note that this approach is the most common default option in many implementations.}
\item Multiple imputation (MI, Section~\ref{sec:mi}): we apply the standard full data estimators from Section~\ref{sec:full-data} on the imputed data (5-10 imputations obtained using the R package \texttt{mice} \citep{mice}) where we use either (a) within-study multiple imputation (WI-MI), (b) ad-hoc multiple imputation (AH-MI), or (c) fixed effect multiple imputation (FE-MI).
\item EM (see Section~\ref{sec:mia-estimators}): we use EM to fit logistic regression models for $r^*(x)=\frac{Pr(i\in Set_{\mathcal{O}}|X_i^*=x^*)}{Pr(i\in Set_{\mathcal{R}}|X_i^*=x^*)}$ and linear regression models for $\mu_{a,trial}^{*}$ on the incomplete data using the R package \texttt{misaem} \citep{jiang_etal_2018}.
\item MIA (see Section~\ref{sec:mia-estimators}): we use generalized random forests with MIA splitting criterion to estimate the generalized regression models $r^*$ and $\mu_{a,trial}^{*}$ on the incomplete data, using the R package \texttt{grf} \citep{athey_etal_AS2019}.
\end{enumerate}

Note that for the EM and MIA approach, we only compute the IPSW, CO and AIPSW estimators (see Section~\ref{sec:full-data-estimators}) since the calibration weighting estimator in its current form is not directly applicable on incomplete data and future work is required to adapt this estimator to incomplete data.

\subsection{Results}

Due to the large number of different scenarios we consider in this simulation study, we first provide an overview in Table~\ref{tab:assumptions} of the different assumption components (e.g., missingness mechanism, causal identifiability) required by the different approaches listed above. This allows to read of the expected behavior of the different estimators in the various cases considered in this experimental study.

\begin{table}[H]
\def\arraystretch{1}
\setlength{\tabcolsep}{3pt}
\centering
\caption{Methods for handling incomplete observations in treatment effect generalization and their assumptions on the underlying data generating process. Throughout all cases, we assume that Assumptions \ref{a:consist}-\ref{a:random} hold. (\cmark indicates cases that can be handled by a method, whereas \xmark \,marks cases where a method is not applicable in theory; (\xmark) indicates cases without theoretical guarantees but with good empirical performance.)}
\label{tab:assumptions}
\begin{tabular}{l|c|c|c||c|c||c|c|}
\cline{2-8}
 \multicolumn{1}{l|}{} &  \multicolumn{3}{c||}{Missingness} & \multicolumn{2}{c||}{\makecell{Identifiability of \\generalized $\tau$}} &  \multicolumn{2}{c|}{\makecell{Models \\for $(S,Y)$}} \\ 
%\cline{2-10} 
\multicolumn{1}{l|}{}   &  \scriptsize MCAR & \scriptsize MAR & \scriptsize MNAR & \scriptsize \makecell{Standard \\($\equiv$ \eqref{a:trans-2} \& \eqref{a:pos})} & \scriptsize \makecell{$CIS$ \\($\equiv$ \eqref{a:trans-2-na}  \& \eqref{a:pos-na})} & \scriptsize \makecell{Generalized\\linear\\models} & \scriptsize \makecell{Non-\\parametric\\models} \\ 

\hline
 \multicolumn{1}{|l|}{\scriptsize \textit{CC}}  & \cmark & \xmark & \xmark & \cmark & \xmark & \cmark  & \cmark  \\ 
\hline
 \multicolumn{1}{|l|}{\scriptsize \textit{EM}} & \cmark & \cmark &\xmark & \xmark & \cmark & \cmark & \xmark \\ 
\hline 
 \multicolumn{1}{|l|}{\scriptsize\textit{MIA}} & \cmark & \cmark &\cmark & \xmark & \cmark & \cmark & \cmark \\ 
\hline
 \multicolumn{1}{|l|}{\scriptsize\textit{MI}} &  \cmark & \cmark & \xmark & \cmark & \xmark & \cmark & (\xmark) \\ 
\hline
\end{tabular}
\end{table}

\begin{figure}[ht!]
\centering
   \includegraphics[width=0.95\textwidth]{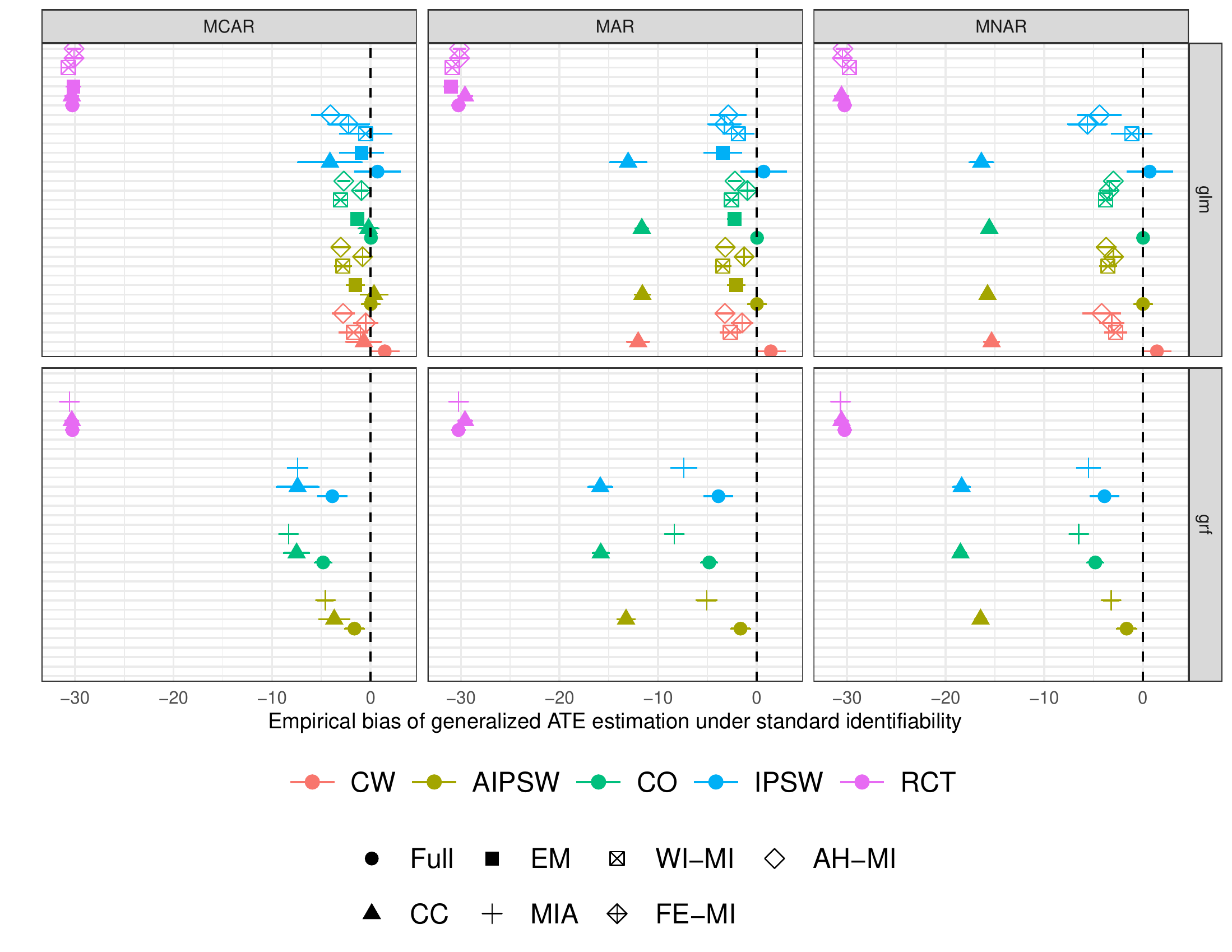}
\caption{Empirical bias of generalizing ATE estimators under the \textit{standard transportability assumption}, 95\% Monte Carlo confidence intervals, $n=1000$. Different colors denote different generalization estimators; different shapes represent different strategies to handle missing values. EM failed to converge in the MNAR scenario and is therefore omitted in the corresponding panel. Full: estimation on full data; CC: estimation on complete cases; EM: estimation by expectation maximization directly handling missing values; MIA: estimation by generalized random forests with missing incorporated in attributes; WI-MI: within-study multiple imputation; FE-MI: fixed effect joint multiple imputation; AH-MI: ad-hoc joint multiple imputation.}
\label{fig:standard-bias-nsim100}
\end{figure}

\begin{figure}[ht!]
    \centering
\includegraphics[width=0.95\textwidth]{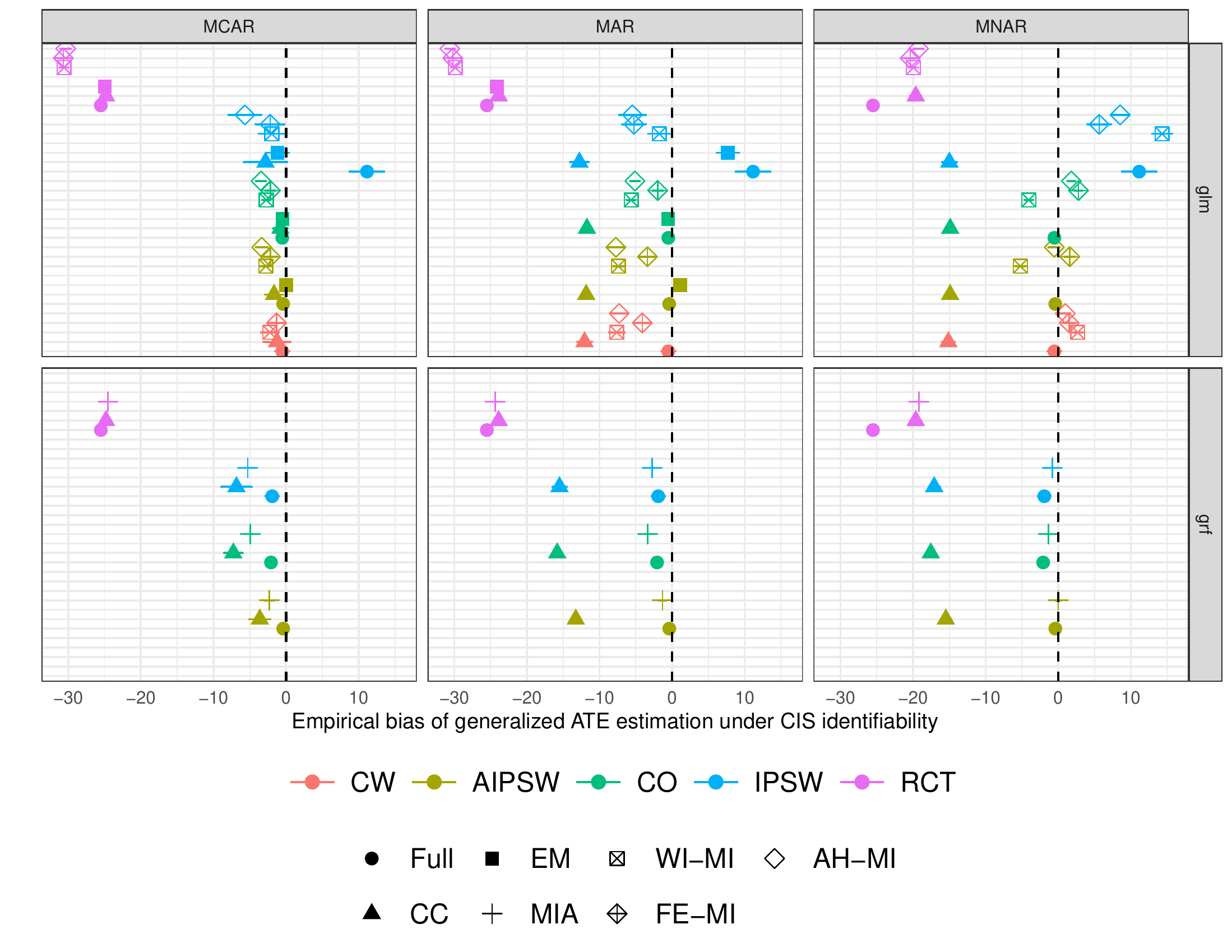}
\caption{Empirical bias of generalizing ATE estimators under the \textit{CIS assumption}, 95\% Monte Carlo confidence intervals, $n=1000$. Different colors denote different generalization estimators; different shapes represent different strategies to handle missing values. EM failed to converge in the MNAR scenario and is therefore omitted in the corresponding panel. Full: estimation on full data; CC: estimation on complete cases; EM: estimation by expectation maximization directly handling missing values; MIA: estimation by generalized random forests with missing incorporated in attributes; WI-MI: within-study multiple imputation; FE-MI: fixed effect joint multiple imputation; AH-MI: ad-hoc joint multiple imputation.}
\label{fig:cis-bias-nsim100}
\end{figure}

Some simulation results are in agreement with what is expected but there are also some gaps between the expected and empirical behavior in the chosen simulation settings. 
Indeed, in Figures~\ref{fig:standard-bias-nsim100} and  \ref{fig:cis-bias-nsim100} we report the empirical bias with at 95\% Monte Carlo confidence interval (based on a Monte Carlo standard error) of the estimated  ATE $\hat{\tau}$ relative to the true value $\tau=27.4$  (or $\tau=58.9$ in the non-linear setting), using $n_{sim}=100$ repetitions of each scenario.
We define the empirical bias and its Monte Carlo standard error respectively by
\begin{equation*}
    \widehat{B}_{\hat{\tau}}=\widehat{Bias}(\hat{\tau}) =\frac{1}{n_{sim}}\sum_{i=1}^{n_{sim}} \hat{\tau}_i-\tau,\quad
    \widehat{SE}(\widehat{B}_{\hat{\tau}}) = \sqrt{\frac{1}{n_{sim}(n_{sim}-1)}\sum_{i=1}^{n_{sim}} (\hat{\tau}_i-\overline{\tau})^2},
\end{equation*}
where $\overline{\tau}=\sum_{i=1}^{n_{sim}} \hat{\tau}_i$, following \citet{morris2019using}.

Based on the findings reported in Figures~\ref{fig:standard-bias-nsim100} and \ref{fig:cis-bias-nsim100} as well as the results in the non-linear setting reported in Appendix~\ref{appendix:results}, we summarize the main observations below.

\subsubsection{Adherence with expected results in the linear case}

\paragraph{Standard identifiability assumptions (Figure~\ref{fig:standard-bias-nsim100}).}
As expected, we observe that the full-data estimations are unbiased in all scenarios under the standard identifiability assumptions \ref{a:trans-2}-\ref{a:pos} in the linear DGP case. The complete case estimations are, unsurprisingly, unbiased only in the MCAR case.
The behavior of the MIA-based estimators is heterogeneous and tends toward biased results for all missingness mechanisms. 
The joint fixed effect MI estimator (FE-MI) comes closer to the expected behavior of the multiple imputation approach than the ad-hoc MI (AH-MI) and the within-study MI (WI-MI) estimator as it has small or no bias under the standard causal identifiability and ignorable missingness assumptions (I+MCAR and I+MAR), but all three fail in the MNAR case (as expected). The EM and MIA based methods are biased in all scenarios which is expected as these estimators rely on the alternative identifiability assumption.

Note that for the full data case, the choice of the estimator, namely parametric (in our case, glm) or non-parametric (here grf), has an impact on the bias, especially for the single-model estimators IPSW and CO. This is not surprising given the linear specification of the selection and outcome models from \eqref{eq:selection-model} and \eqref{eq:outcome-model} and the rather slow convergence of the chosen non-parametric method, random forest (\texttt{grf}), for linear models.\footnote{The random forest approach would require a lot of data to estimate linear regression functions; random forests are however known for their good performance in the presence of non-linearities and high order interaction terms \citep{breiman2001random}.}

\paragraph{Alternative identifiability assumptions (Figure~\ref{fig:cis-bias-nsim100}).}
Under the CIS assumption, only the full-data estimators that (partly) rely on the outcome model, namely CO, AIPSW, and CW are unbiased, whereas the parametric IPSW estimator fails under CIS. This is expected as the selection model depends on both the observed values and the missingness pattern, the latter not being available in the full-data case; on the other hand, in the simulated data, the conditional outcome model does not depend on the missingness pattern.
The behavior of the EM estimations is as expected: all estimators are unbiased under CIS under MCAR and MAR. However the AIPSW estimator performs better than the IPSW and the CO. In the MNAR case, the EM algorithm fails to converge. Under the non-linear case, this linear approach fails to recover the true ATE in all cases of missing values mechanisms.
The MIA estimations overall have either small or no bias under CIS, especially the AIPSW estimator. Furthermore, under the CIS assumption, the AIPSW estimator always performs at least as well as the simple estimators (IPSW and CO). The good performance of MIA in the MNAR case is not surprising as this method can handle the MNAR case by definition \citep{josse_etal_2019}.
 
 \subsubsection{Deviations from expected results}
 \paragraph{Standard identifiability assumptions (Figure~\ref{fig:standard-bias-nsim100}).}
Surprisingly, the  MI IPSW estimators are biased in all cases except the within-study MI estimator in the MCAR and MNAR case. This behavior is not expected and is likely due to low sample size and small number of imputations compared to the multiple steps of this estimator (imputation model and nuisance parameter estimation) which cannot compensate for each other's inaccuracies. In additional simulations the bias decreases as the sample size and number of multiple imputations increases (not reported).
 
 \paragraph{Alternative identifiability assumptions (Figure~\ref{fig:cis-bias-nsim100}).}
 Surprisingly, the parametric full-data IPSW estimator recovers the true value in the MAR and MNAR case under the alternative assumptions \ref{a:trans-2-na}-\ref{a:pos-na}. One could attempt to explain this through the fact that missing values can be explained based on the observed values in the MAR case and based on the full data (effectively observed and missing values) in the MNAR case, thus the response pattern could be modeled and recovered from the full data, and therefore the selection model becomes identifiable from the full data.

\subsubsection{Comments on the non-linear setting.}

The detailed results of the simulations in the non-linear case, i.e., the simulations based on the non-linear selection and outcome models \eqref{eq:selection-model-nonlin} and \eqref{eq:outcome-model-nonlin}, are presented and commented in Appendix~\ref{appendix:results}.

\subsubsection{Conclusion from simulation study}

In view of the results, if one has to recommend a method, it is preferable, from the present empirical evidence, to choose the MIA-AIPSW estimator or the joint multiple imputation coupled with the CW estimator (FE-MI + CW). Both approaches are simple to use but can be computationally costly and may require large sample sizes if the true data generating process is very non-linear, i.e., not approximately linear. Additionally, the performance of a chosen estimator might depend on the true underlying missingness mechanism, e.g., the multiple imputation based estimators generally assume a MAR mechanism. 
In the simulation study, certain estimators performed better than expected from the theoretical considerations; a closer examination of these cases might help refining the set of sufficient identifiability assumptions.

We conclude this simulation study with remark on inference; we have focused on bias of the proposed methods and less on inference and how well uncertainty is captured by the confidence intervals. In a preliminary simulation study, we find that a non-parametric bootstrap yields slightly lower coverage than the nominal level.\footnote{For a nominal level of 95\%, with fixed effect MI we obtain coverage of between 84\% (IPSW) and 89\% (CW) in a linear setting (\eqref{eq:selection-model} and \eqref{eq:outcome-model}) with missing values from a MAR mechanism.} A systematic work addressing this issue is however still required, even for task of treatment effect estimation from a single data source as works on this topic are currently still scarce \citep[see, e.g.,][]{imbens2018causal, little2019causal}.

\subsection{Impact of different proportions of missing values in the RCT and observational data}\label{sec:diff-propNA}

\begin{figure}[ht!]
    \centering
    \includegraphics[width=0.7\textwidth]{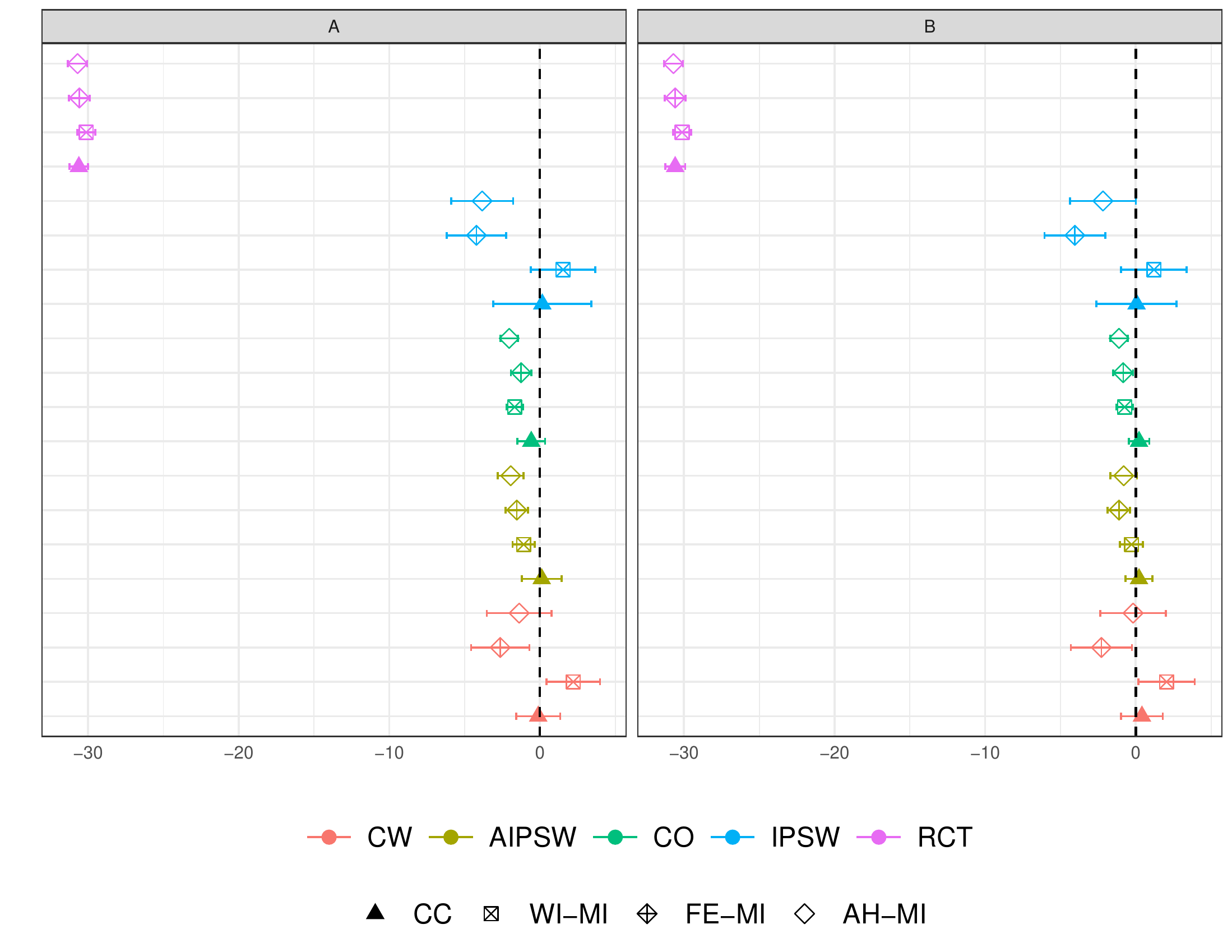}
\caption{Bias estimates of generalized ATE under \textit{standard transportability assumption \ref{a:trans-2}} where missing values are \textbf{``study-wise MCAR''}. For $n=1000$, 100 repetitions,  95\% Monte Carlo confidence intervals. Case A=\{m=$10\times n$, RCT=10\% NA, Obs=50\% NA\}; case B=\{$m=10\times n$, RCT=5\% NA, Obs=22\% NA\}. Different colors denote different generalization estimators; different shapes represent different strategies to handle missing values. CC: estimation on complete cases; WI-MI: within-study multiple imputation; FE-MI: fixed effect joint multiple imputation; AH-MI: ad-hoc joint multiple imputation.}
\label{fig:Standard-MCAR-diffpropNA}
\end{figure}

It is common that the RCT presents significantly less missing values than the observational study due to a more systematic monitoring of the data collection process. This invokes the question of how the above studied methods behave in the case of unbalanced proportions of missing values or different missing values mechanisms in the different data sources. 
Extending the previous simulation study by such a case, we summarize in Figure~\ref{fig:Standard-MCAR-diffpropNA} the performance of the different estimators under different scenarios of varying proportions of missing values in the RCT and the observational data when the data are MAR given the source indicator (or equivalently, we say it is MCAR in each data set).

As expected, the complete case estimators are unbiased in this special case since conditionally on the source indicator, the data are MCAR. For the other estimators, results are similar across different proportions of missing values, either 10\%, 50\% in RCT and observational data respectively, or 5\% and 22\% respectively. 
These results are supporting our claim that the previous results and methods apply as well to the likely case of different proportions of missing values in the two studies. Indeed the following data analysis in Section~\ref{sec:data-analysis} is an example of this case.

\section{Application on critical care data}\label{sec:data-analysis}

In this part, we come back to the medical question introduced in the beginning of this work about the potential effect of tranexamic acid (TXA) on mortality in patients with intracranial bleeding admitted in French trauma centers, which we consider as representatives of the core entities in western European major trauma management infrastructure who admit an increasing proportion of elderly major trauma patients  \citep{roozenbeek2013changing}. 
We recall that, in order to answer this question, we have at disposal two data sources: (1) CRASH-2, a multi-center international RCT, (2) Traumabase, an observational national registry.
A detailed data analysis of the observational registry to address the above medical question has been conducted by \citet{Mayer2020}. We thus refer to this previous analysis for a detailed description of the observational registry as well as their findings. These can be summarized as follows: leveraging only the observational registry does not provide evidence towards a beneficial (or detrimental) effect of TXA on trauma patients with TBI in terms of head-injury-related mortality.

We will first recall a summary of the findings of the original CRASH-2 study \citep{crash2} before turning to focus on how the handling of missing values in the RCT and the observational registry impacts the final estimations of the population average treatment effect.

\subsection{Findings of the CRASH-2 RCT}

The CRASH-2 (Clinical Randomisation of Antifibrinolytic in Significant Haemorrhage) trial enrolled 20,211 patients in 274 hospitals in 40 countries between May 2005 and 2009 \citep{crash2}.\footnote{Note that none of the centers participating in the CRASH-2 trial was located in France, while Traumabase registry only contains patients admitted in French trauma centers. However, several western European countries were among the 40 participating countries of the CRASH-2 trial.}
The aim of this trial was to study the  effect of tranexamic acid in adult trauma patients with ongoing significant hemorrhage or at risk of significant hemorrhage, within 8 hours of injury (inclusion criteria), except those for whom antifibrinolytic agents were thought to be clearly indicated or clearly counter-indicated (exclusion criteria)\footnote{Extract from the study protocol available at \url{https://www.thelancet.com/protocol-reviews/05PRT-1}.}.
More precisely, eligible patients were defined as trauma patients within 8 hours of the injury, of age at least 16 years (i) with ongoing significant hemorrhage (systolic blood pressure less than 90 mmHg and/or heart rate more than 110 beats per minute), or (ii) who are considered to be at risk of significant hemorrhage. 
The inclusion criteria and other baseline regressors are summarized in a causal diagram in Appendix~\ref{appendix:analysis} (Figure~\ref{fig:crash2-txa-dag}).

The results of the CRASH-2 study are reported in \citet{crash2} and show a beneficial effect of TXA on the trial population for the primary outcome of interest (all-cause 28 day death).

\subsection{Integration of the CRASH-2 trial and the Traumabase registry}

In the following, we discuss common variables definition, outcome, treatment, and designs in order to leverage both sources of information. We recall the causal question of interest: ``What is the effect of the TXA on brain-injury death on patients suffering from TBI?'' This part is important for the harmonization of the study protocol.

\begin{itemize}
\item \textbf{Treatment exposure.}
The treatment protocol of CRASH-2 frames the timing and mean of administration precisely (a first dose given by intravenous injection shortly after randomization, i.e., within 8 hours of the accident, and a maintenance dose given afterwards \citep{crash2}). 
In the Traumabase study, which is a retrospective analysis, this level of granularity concerning TXA is unfortunately not available. Neither the exact timing, nor the type of administration are specified for patients who received the drug. However, the expert committee agreed that the assumption of treatment within 3 hours of the accident is very likely since this drug is administered in pre-hospital phase or within the first 30 minutes at the hospital \citep{Mayer2020}.
\item \textbf{Outcome of interest.}
The CRASH-2 trial defined primary outcome as any-cause death in hospital within 28 days of injury. This outcome is also available in the Traumabase.
\item \textbf{Covariates accounting for trial eligibility.}
As noted in Section~\ref{sec:alternative}, the distributional shift between trial and target population can be seen from the selection perspective as well where the shifted variables are those related to the eligibility (and participation) criteria. For the CRASH-2 trial, four criteria determined inclusion: age (patients of at least 16 years old were eligible), ongoing or risk of significant hemorrhage (defined as systolic blood pressure below 90 mmHg or heart rate above 110 beats per minute, or clinicians evaluation of a risk), within 8 hours of injury and absence of a clear indication or counter-indication of antifibrinolytic agents. The necessary variables are also available in the Traumabase, either exactly or in form of proxies, which allows the estimation of the trial inclusion model on the combined data. 
With these eligibility criteria, we obtain a sample size of 8248 patients in the Traumabase and of 3727 patients in the CRASH-2 trial.\footnote{Except for under-aged patients and patients with isolated TBI, all registry patients were potentially trial-eligible because of the rather broad inclusion criterion of ``risk of significant hemorrhage'' which is difficult to assess without medical imaging tools in the first few hours after an accident \citep{hamada2018development}.}
\item \textbf{Additional covariates.}
Note that other covariates are (partially) available in both data sets, while not responsible of trial inclusion according to CRASH-2 investigators. But as this could still be covariates moderating the outcome and treatment effect, we include them in the outcome regression  models used in the CO and AIPSW estimators \eqref{eq:g-formula} and \eqref{eq:aipsw} to improve precision of the estimators. Based on expert knowledge and the available information from both data sources, we include three additional variables in the analysis: sex (binary), type of injury (categorical, $1=$blunt, $2=$ penetrating, $3=$ blunt and penetrating), and initial Glasgow coma scale (numeric, integers from $3$ to $15$). Note that these three covariates are all mentioned in the baseline of CRASH-2 results \citep{crash2}, arguing that they should impact the outcome. The variables \textit{central capillary refill time} and \textit{respiratory rate} are also mentioned but are not available in the Traumabase, we thus omit them from this joint study.
\item \textbf{Missing values.}
First, note that the RCT contains almost no missing values, whereas the variables for determining eligibility in the observational data contains important fractions of missing values, as shown in Table~\ref{fig:percentagenaeligibility}. 
\begin{figure}[ht!]
    \centering
    \includegraphics[width=0.6\textwidth]{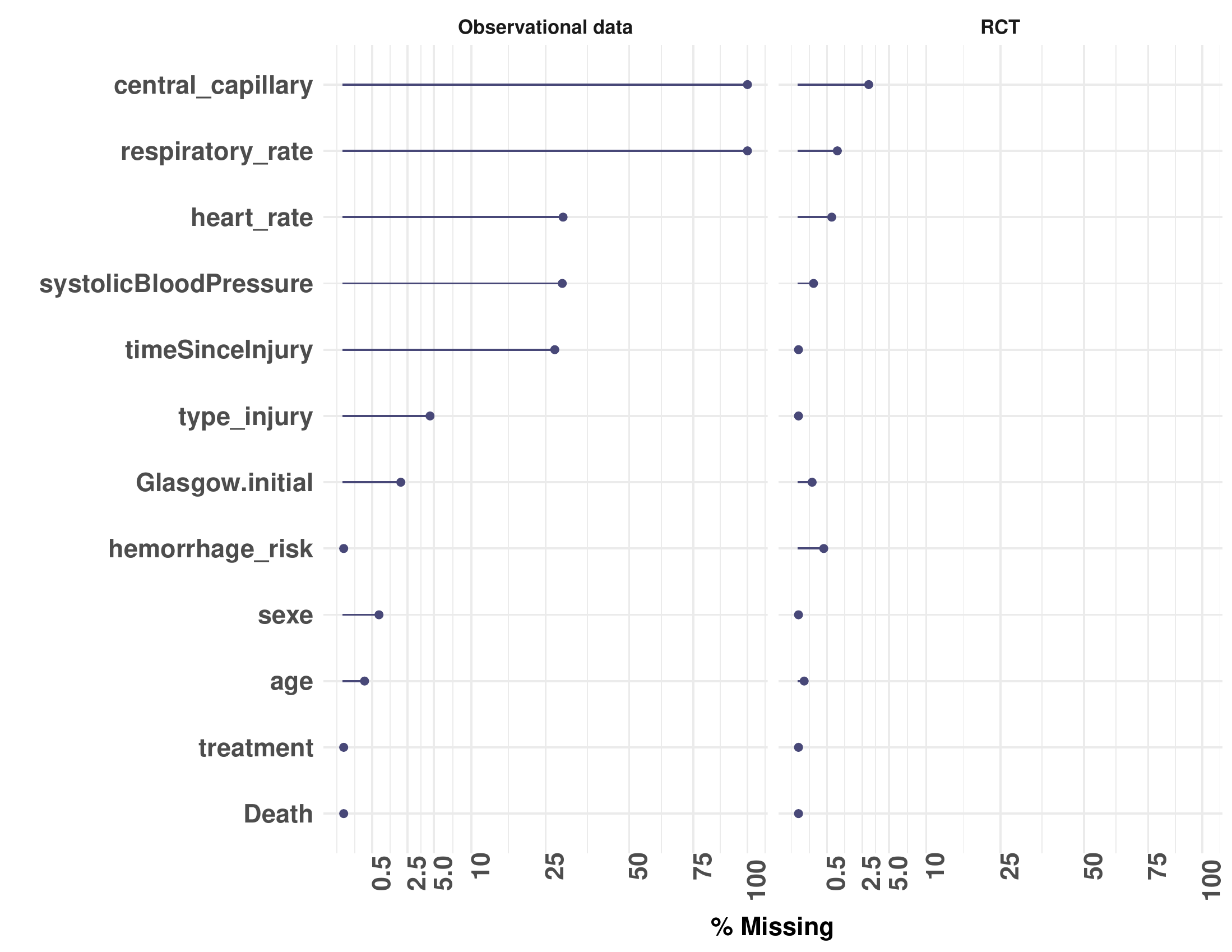}
    \caption{Percentages of missing values in each covariate for the Traumabase and CRASH-2 RCT.}
\label{fig:percentagenaeligibility}
\end{figure}
While the MCAR assumption is reasonable for the RCT \citep{crash2}, the missing values in the observational Traumabase are more complex and, according to the medical experts monitoring the collection process, partly non-ignorable. For example, the pre-hospital systolic blood pressure (SBP) is likely to be missing for patients with severe ongoing bleeding. Since the latter is informed in the \textit{hemorrhage risk} variable, we could admit the missing values in the SBP variable as being MAR. A similar reasoning can be applied for the delay between the accident and treatment administration. However, there remains uncertainty as to whether the observed variables allow to fully explain the missingness in this variable.
\item \textbf{Distribution shift.}
There are different ways of assessing the shift between the distributions of the two studies, e.g., by univariate comparisons. When comparing group means, we note that the average patient age in the RCT is 7-9 years below the average age in the observational study; since older patients with head injuries frequently have comorbidities that increase the risk of infection, surgical risk and mortality after the injury, this difference in age is relevant in this study given the primary endpoint of in-hospital mortality \citep{wutzler2009association}. In Appendix~\ref{appendix:analysis}, we provide a simplified comparison of the means of the covariates between the treatment groups of the two studies in Figure~\ref{fig:catdes_eligibility}.
\item \textbf{Transportability of the CATE. }
Due to the design of the CRASH-2 study, namely the eligibility criteria which all need to be observed to decide upon trial eligibility, the alternative transportability assumption CIS \ref{a:trans-2-na} is less plausible to hold in this case and we rather consider the standard causal identifiability assumptions \ref{a:trans-2} and \ref{a:pos} to be satisfied by the CRASH-2 and Traumabase studies. We consider that the assumptions for the multiple imputation strategy, namely the MCAR or MAR mechanism \eqref{eq:mcar} and \eqref{eq:mar} (and \eqref{eq:mcar-target} and \eqref{eq:mar-target}), can be considered plausible in this real-world example.
The details of the assessment of the support inclusion assumptions \ref{a:pos} and \ref{a:pos-na} can be found in Appendix \ref{appendix:analysis}. In summary, we find that the selection scores obtained using EM and MI are similar and may suggest that the support inclusion assumption is satisfied since we observe a good degree of overlap between the distributions of the scores for the two data sets. The scores estimated via MIA however concentrate around 0 and 1 for the observational and RCT observations respectively, apparently contradicting the initial assumption of overlap between the RCT and observational study populations. This apparent contradiction with expert knowledge about the respective populations is rather an indicator that the MIA approach and its underlying assumption, namely the CIS assumption \ref{a:trans-2-na} (which implies that the distributional shift depends on the observed values and the missingness pattern) are not appropriate in this study. 
These results provide an additional argument in favor of the multiple imputation strategy which appears to be more adapted to the handling of missing values in this analysis; in particular we will apply a joint fixed effect multiple imputation strategy since it outperforms the other multiple imputation strategies in the simulation study of Section~\ref{sec:simulations}.
\end{itemize}

\subsection{Results of generalized ATE from CRASH-2 to the observational target population}

We now apply the estimators presented in this work and implemented first for the simulation study of Section~\ref{sec:simulations}. The confidence intervals for the corresponding point estimators are computed via non-parametric stratified bootstrap \citep{Efron1994} using 100 bootstrap samples (using stratified sampling to preserve the study-specific sample sizes).
We additionally report two consistent ATE estimators from the solely CRASH-2 data: the difference in mean estimator (\texttt{Difference in mean}) and the difference in conditional means where  we assume logistic regression models of the outcome on covariates for the treated and controls.   (\texttt{Difference in conditional mean}).
The former only involves treatment assignment $A$ and outcome $Y$ and thus requires no additional handling of the incomplete covariates; the latter is obtained using an EM algorithm for logistic regression with ignorable missing values in the covariates \citep{jiang_etal_2018}.
We also recall the results of the generalized random forest based AIPW estimators \citep{wager_athey_JASA2018} for the observational study applied solely on the Traumabase data (see \citet{Mayer2020}).
Since AIPW combined with either missing incorporated in attributes (\texttt{MIA AIPW}) or multiple imputation (\texttt{MI AIPW}) is recommended by \citet{Mayer2020} when analyzing observational data, these are the estimators reported in this analysis.

\begin{figure}[ht!]
    \centering
    \includegraphics[width=0.9\textwidth]{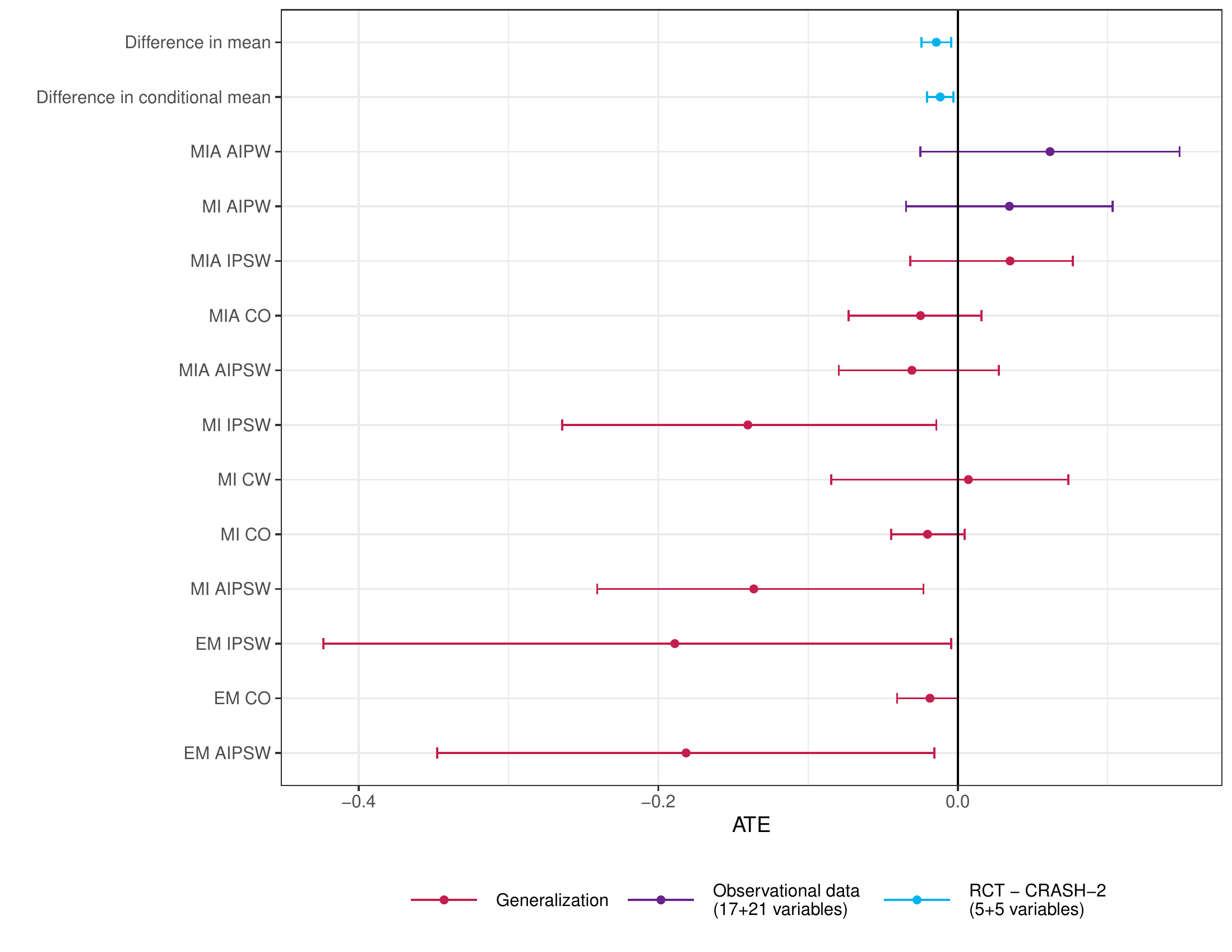}
    \caption{Separate and joint ATE estimators and 95\% confidence intervals computed on the Traumabase (observational data set; purple), on the CRASH-2 trial (RCT; cyan), and generalized from CRASH-2 to the Traumabase target population (red). Number of variables used for adjustment in each context is given in the legend. The confidence intervals obtained on the observational data set and on the joint data sets are obtained via nonparametric bootstrap.}
    \label{fig:results_crash2_traumabase}
\end{figure}

When summarizing the results from the separate analyses on the RCT and the observational data respectively and the results from the joint analysis of both studies, we observe on Figure~\ref{fig:results_crash2_traumabase} a discrepancy between the different results.\footnote{With the exception of the calibration weighting estimator, all MI-based estimators are followed by GRF-based regressions for the components of the IPSW, CO and AIPSW estimators. This choice is motivated by expert knowledge of the high complexity of the data and underlying mechanisms of the treatment and of trauma-related injuries which are believed to be better captured by non-parametric methods such as random forests.}
However, in view of the simulation study and the domain knowledge of our clinical experts, we have more reason to believe in the results of methods based on multiple imputation.
The high variability of the IPSW results is most probably due to extreme weights which are common in practice, even with stabilized weights. This is why in this case, and in general, we tend to have more confidence in estimations from AIPSW and CW.
Finally, the heterogeneity in these findings aligns with the results from the simulation study: in practice, results depend on how the covariate shift and missingness mechanisms are dealt with (in identifiability assumptions and in estimation).

The large confidence intervals could be partly explained by the measurement noise in the administration delay variable in the Traumabase\footnote{Note that we use stabilized weights instead of the standard weights.} Another direction to explore could be to trim the scores such as to avoid extreme weights. This could potentially reduce the size of the confidence intervals but it is known to generally provide a biased result since trimming of the weights induces an implicit change in the definition of the target population \citep{li_etal_JASA2018}: contrary to the RCT, the Traumabase does not encode the exact delay of treatment administration, but is defined by a noisy proxy (delay between accident and admission to the resuscitation bay). However there exists evidence that administration delay is a treatment modifier for TXA and that only early administration has a beneficial effect \citep{Hijazi2015, Crash2011}. This remark and the discrepancies between the findings, especially the different conclusions of the joint fixed effect multiple imputation estimators call for additional attempts to further refine the administration delay proxy variable in the Traumabase and potentially for additional analyses with supplementary data such as the CRASH-3 study \citep{crash3} which describes another slightly different TBI patient population.

\section{Conclusion}

In this work we have proposed several estimators that are suited for generalizing a treatment effect from an incomplete RCT to a different target population described by an incomplete observational covariate data. One of the difficulty was to account for the different variables in both sources. Which of the proposed methods is preferable depends on the underlying identifiability assumptions for generalizing the treatment effect. Under the standard full data causal identifiability assumptions in combination with MCAR and MAR assumptions on the missingness mechanism, we recommend a joint multiple imputation that models the data source as fixed effect as long as the missing values are ignorable. 
If the identifiability assumptions are altered to account for informative missing values implicated in the selection process, then estimators involving generalized conditional regression models are suited. These are however strong assumptions and 
 in many common examples in medicine or epidemiology these do not appear to be the most plausible ones. Nevertheless, in certain contexts such as in pragmatic trials with minimal sets of eligibility criteria these assumptions could potentially be relevant.

On the methodological side, for all considered approaches and simulation settings, the question of varying missing values mechanisms across data sources remains to be addressed in more detail. Note as well that we have focused on missing values in both data sets, but the recommendations extend to the case where there are only missing values in the observational data and not the RCT due to different levels of systematic data collection.
The problem of incomplete observations addressed here is different from the problem of inconsistent variable sets between an RCT and an observational dataset, e.g., one variable is completely missing in one set, which is more related to unobserved confounding and a recent work proposes sensitivity analysis to address this issue \citep{colnet_etal_2021}.

Finally, it should be noted that, contrary to what is often perceived by practitioners, large proportions of missing values are not an obstacle to statistical analysis per se, but the task of understanding the role of their generating process and of combining this with the original full data assumptions is challenging, as are the adjustments to subsequent estimators. It is crucial to think carefully about the plausibility of the assumptions, because on our examples the considered approaches give different results.  We therefore think an important direction of future work could focus on establishing sensitivity analysis bounds to address missing values in this context.

\paragraph{Acknowledgement}
We would like to thank Shu Yang for fruitful discussions and her valuable feedback on our work. We thank Tobias Gauss, Jean-Denis Moyer and Fran\c{c}ois-Xavier Ageron for their medical insights and interpretation of our data analysis; we thank the CRASH-2 trial investigators for sharing the trial data with us. Finally, we would like to thank the anonymous reviewers whose detailed feedback helped us improve this work. This project was done in part while IM and JJ were visiting the Simons Institute for the Theory of Computing.\\
\noindent {\bf{Funding}} IM was supported by a EHESS PhD fellowship and a Google PhD fellowship.
The funding institutions had no role in the design and conduct of the study; collection,
management, analysis, and interpretation of the data; preparation, review, or approval of the
manuscript; and decision to submit the manuscript for publication.

\vspace*{1pc}

\noindent {\bf{Conflict of Interest}}

\noindent {\it{The authors have declared no conflict of interest.}}

\newpage

\appendix

\section*{Appendix}

\section{Details on the background full data identification}\label{appendix:full-data}

\subsection{Identification formula}
Under Assumptions \ref{a:consist} - \ref{a:pos} 
the ATE can be identified
based on the following formulas: 
\begin{enumerate}
\item Reweighting formulation:
\begin{align*}
    \tau &= \mathbb{E}_{target}\left[\tau_{target}(X)\right]&&\text{Law of total expectation}\\
    &= \mathbb{E}_{target}\left[\tau_{trial}(X)\right]&&\text{Assump. \ref{a:trans-2}}\\
    & = \mathbb{E}_{target}\left[\mathbb{E}_{trial}\left[ \left( \frac{A}{e_{trial}(X)} - \frac{1-A}{1-e_{trial}(X)} \right) Y| X\right]\right] &&\text{Identifiability of } \tau_{trial}(X)\\
    & = \mathbb{E}_{trial}\left[r(X) \left( \frac{A}{e_{trial}(X)} - \frac{1-A}{1-e_{trial}(X)} \right) Y\right]&&\text{Assump. \ref{a:pos}.}\,, \label{eq:idweight}
\end{align*}
where $\tau_{trial}(x) = \mathbb{E}_{trial}[Y(1) - Y(0) | X=x]$ is identifiable under assumptions \ref{a:consist} and \ref{a:random} and $e_{trial}(x)=Pr(A=1|X=x)$ is the treatment propensity.
Note that the density ratio cannot directly be estimated from the data, but it requires estimating conditional odds: the conditional odds $\alpha(x)$ can be estimated by fitting a regression model that discriminates RCT versus observational samples, and Bayes' rule gives:
\begin{equation*}
    \begin{split}
\alpha(x) 
&= \frac{Pr(i \in Set_{\mathcal{R}}\,|\,  X_i=x)}{1-Pr(i \in Set_{\mathcal{R}} \,|\,  X_i=x)} \\
&= \frac{Pr(i \in Set_{\mathcal{R}})}{Pr(i \in Set_{\mathcal{O}})}\times \frac{Pr(X_i = x \,|\, i \in Set_{\mathcal{R}})}{Pr(X_i = x \,|\, i \in Set_{\mathcal{O}})}\\
&= \frac{Pr(i \in Set_{\mathcal{R}})}{Pr(i \in Set_{\mathcal{O}})}\frac{p_{trial}(X)}{p_{target}(X)} \\
& \,\,\approx \frac{n}{m} \frac{p_{trial}(X)}{p_{target}(X)} = \frac{n}{m} \frac{1}{r(X)}.
\end{split}
\end{equation*}
Finally we have
$$
\tau = \mathbb{E}_{trial}\left[r(X) \tau_{trial}(X) \right]\,.
$$
Note that in a fully randomized RCT where $e_1(x)=0.5$ for all $x$, this formula further simplifies to
$$
\tau = \mathbb{E}_{trial}\left[2r(X)(2A-1)Y\right]\,.
$$
\item Regression formulation:
\begin{equation*} 
\label{eq:idereg}
\tau=\mathbb{E}_{target}[\mu_{1,trial}(X)-\mu_{0,trial}(X)] \,,
\end{equation*} 
which can be shown using the following derivations for $a\in\{0,1\}$.
  \begin{align*}
    \mathbb{E}_{target}[Y(a)] &=  \mathbb{E}_{target}\left[ \mathbb{E}_{target}[Y(a) \mid X] \right] &&\text{Law of total expectation}\\
    &=\mathbb{E}_{target}\left[ \mathbb{E}_{trial}[Y(a) \mid X] \right] &&\text{Assump. \ref{a:trans-2}} \\
    &= \mathbb{E}_{target}\left[ \mathbb{E}_{trial}[Y(a) \mid X, A = a] \right] &&\text{Assump. \ref{a:trans-2}}\\
    &= \mathbb{E}_{target}\left[ \mathbb{E}_{trial}[Y \mid X,  A = a] \right] &&\text{Assump. \ref{a:consist}}\\
    & = \mathbb{E}_{target}[\mu_{a,trial}(X)] &&
  \end{align*}

\end{enumerate}

\section{Details on the incomplete data problem}\label{appendix:missing-data}

\begin{figure}[ht!]
\scriptsize
\setlength{\tabcolsep}{2.7pt}
\begin{center}
\begin{tabular}{|c|c|ccc|c|c|c|} 
 \hline
\multirow{2}{*}{$i$} &  \multirow{2}{*}{Set} & \multicolumn{3}{c|}{Covariates} & Treatment & \makecell{Outcome \\under A=0} & \makecell{Outcome\\under A=1} \\
 &  &$X_1$ & $X_2$ & $X_3$ & $A$ & $Y(0)$ & $Y(1)$ \\ 
 \hline
1 & $Set_{\mathcal{R}}$  &1.1 & 20 & 5.4 & 1 &  23.4 & 24.1\\ 
$\dots$ & $Set_{\mathcal{R}}$   & & $\dots$ & & $\dots$ & $\dots$  & $\dots$ \\
$n-1$ & $Set_{\mathcal{R}}$  &-6 & 45 & 8.3 &  0 & 26.3 & 27.6 \\ 
$n$& $Set_{\mathcal{R}}$  &0 & 15 & 6.2 & 1 & 28.1 & 23.5 \\ 
$n+1$& $Set_{\mathcal{O}}$  & -2 & 52 & 7.1 & \texttt{NA} & \texttt{NA} & \texttt{NA} \\
$n+2$ & $Set_{\mathcal{O}}$  &-1 & 35 & 2.4 & \texttt{NA} &    \texttt{NA} & \texttt{NA}\\
$\dots$ & $Set_{\mathcal{O}}$   & & $\dots$ & & \texttt{NA} & \texttt{NA}  & \texttt{NA} \\
$n+m$& $Set_{\mathcal{O}}$  &-2 & 22 & 3.4 & \texttt{NA} &\texttt{NA} & \texttt{NA}\\
 \hline
\end{tabular}
 \hfill
\begin{tabular}{ |c|c|ccc|c|c| } 
 \hline
\multirow{2}{*}{$i$} & \multirow{2}{*}{Set} &\multicolumn{3}{c|}{\makecell{Covariates\\}} & \makecell{Treatment\\} & \makecell{Outcome \\under A} \\
&  & $X_1^*$ & $X_2^*$ & $X_3^*$ & $A$ & $Y$ \\ 
 \hline
1& $Set_{\mathcal{R}}$  &1.1 & 20 & \texttt{NA} & 1 &  24.1 \\ 
$\dots$& $Set_{\mathcal{R}}$  & & $\dots$ & & $\dots$ & $\dots$   \\
$n-1$ &  $Set_{\mathcal{R}}$  &-6 & \texttt{NA} & 8.3 & 0  & 26.3  \\ 
$n$&  $Set_{\mathcal{R}}$ &0 & 15 & 6.2 & 1 & 23.5  \\ 
$n+1$&  $Set_{\mathcal{O}}$  &-2 & 52 & \texttt{NA} & \texttt{NA} & \texttt{NA}  \\
$n+2$ &  $Set_{\mathcal{O}}$  &-1 &  \texttt{NA} & 2.4 & \texttt{NA} & \texttt{NA} \\
$\dots$& $Set_{\mathcal{O}}$ & & $\dots$ & & \texttt{NA} & \texttt{NA}   \\
$n+m$& $Set_{\mathcal{O}}$ &  \texttt{NA} &  \texttt{NA} &3.4 & \texttt{NA} & \texttt{NA} \\
 \hline
\end{tabular}
\end{center}
\caption{Example of data structure in the incomplete data problem setting. Left: complete but never observed underlying data. Right: observed incomplete data.}
\label{fig:typicalsituation-na}
\end{figure}

\section{Details on the estimation methods with missing values}\label{appendix:methods}

\subsection{Schematic illustration of different multiple imputation strategies}

\begin{figure}[ht!]
\centering
    \includegraphics[width=0.7\textwidth]{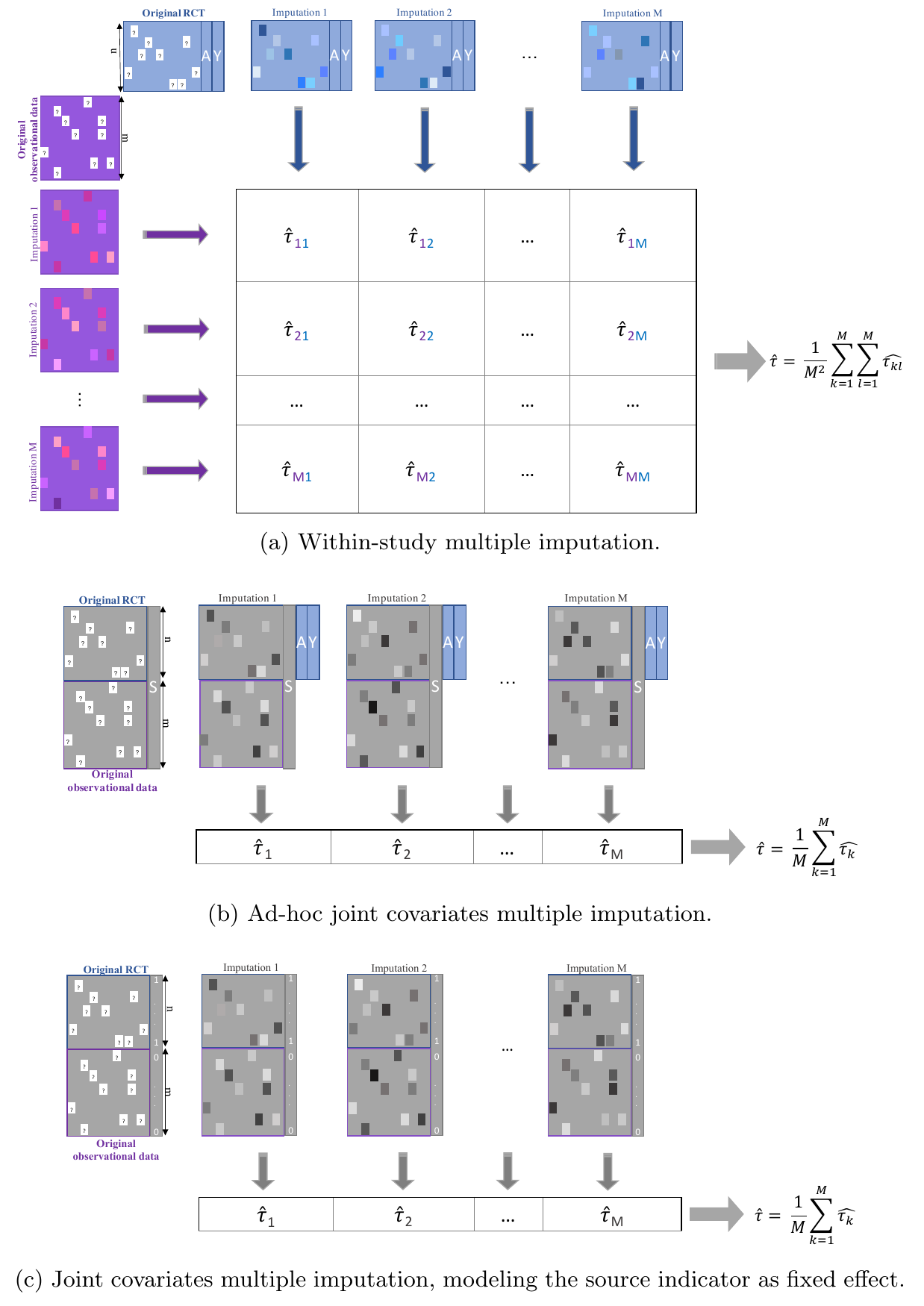}
\caption{Schematic illustrations of different multiple imputation strategies.}
\label{fig:mi}
\end{figure}

\subsection{Prediction on new incomplete observations with parametric model} 

As mentioned in Section~\ref{sec:mia}, it is possible to predict the outcome $y$ for new incomplete observations, using the regression model estimated via EM, by marginalizing over the distribution of missing data given the observed.
More formally, in the logistic regression case, using a Monte Carlo approach and \textit{maximum a posteriori} estimator, it is possible to predict the response $y$ for a new observation $x_i$ as follows:

\begin{enumerate}
\item Sample $$
\left(x_{\mathrm{mis}}^{(k)}, 1 \leq k \leq K\right) \sim \mathrm{p}\left(x_{\mathrm{mis}} \mid x_{\mathrm{obs}}\right)
$$
\item Predict the response $y$ by maximum a posteriori
$$
\begin{aligned}
\hat{y}=\underset{y}{\arg \max } \mathrm{p}\left(y \mid x_{\mathrm{obs}}\right) &=\underset{y}{\arg \max } \int \mathrm{p}(y \mid x) \mathrm{p}\left(x_{\mathrm{mis}} \mid x_{\mathrm{obs}}\right) d x_{\mathrm{mis}} \\
&=\underset{y}{\arg \max } \mathbb{E}_{p_{z_{\operatorname{mis}} \mid x_{o b s}}} \mathrm{p}(y \mid x) \\
&=\underset{y}{\arg \max } \sum_{k=1}^{K} \mathrm{p}\left(y \mid x_{\mathrm{obs}}, x_{\mathrm{mis}}^{(k)}\right)
\end{aligned}
$$
\end{enumerate}

For the linear case, the prediction proceeds in two steps:
\begin{enumerate}
    \item Imputation of the new observation using the estimated variance-covariance matrix of the covariates $\widehat{\Sigma}$.
    $$\hat{x}^{new}_{mis} = -\left[\widehat{\Sigma^{-1}}_{mis,mis}\right]^{-1}\widehat{\Sigma^{-1}}_{mis,obs}x^{new}_{obs}$$
    \item Prediction of response $y$ using the imputed observation $[x^{new}_{obs},\,\hat{x}^{new}_{mis}]$.
\end{enumerate}

\section{Details on the simulation study}\label{appendix:results}

\subsection{Missingness mechanisms}

In our simulation study, we consider all three classes of missingness mechanisms as defined by Rubin's taxonomy.

\begin{enumerate}
    \item MCAR where the probability to have missing values does not depend on any variable: 
        \begin{align}
        \begin{split}
            \label{eq:mcar-simulation-mi}
            \text{(in the trial)} &\quad P(R_{i\cdot}=r|X_i, Y_i, A_i) = P(R_{i\cdot}=r)\\
            \text{(in the target)} &\quad P(R_{i\cdot}=r|X_i) = P(R_{i\cdot}=r)
        \end{split}
        \end{align}
    \item MAR: 
        \begin{align}
        \begin{split}
        \label{eq:mar-simulation-mi}
        \text{(in the trial)} &\quad P(R_{i\cdot}=r|X_i, Y_i, A_i) = f_{trial}(X_{obs(r)}, A_i, Y_i),\\
        \text{(in the target)} &\quad P(R_{i\cdot}=r|X_i) = f_{target}(X_{obs(r)}),\\
        \end{split}
        \end{align}
        for some functions $f_{trial}: \mathcal{X}\times\{0,1\}\times \mathbb{R} \rightarrow [0,1],$  and $f_{target}: \mathcal{X} \rightarrow [0,1],$.
        For example, missing values in $X_1$ are introduced for observation $i$ using a logistic model on $X_2, X_3, X_4$, assuming these three variables are observed for observation $i$.
    \item MNAR: 
        \begin{align}
        \begin{split}\label{eq:mnar-simulation-mi}
        \text{(in the trial)} &\quad P(R_{ij}=0|X_i, Y_i, A_i) = g_{trial}(X_{ij}, Y_i, A_i),\\
        \text{(in the target)} &\quad P(R_{ij}=0|X_i) = g_{target}(X_{ij}),\\
        \end{split}
        \end{align}
        for some functions $g_{trial}: \mathcal{X}_j\times\{0,1\}\times\mathbb{R} \rightarrow [0,1],$ and $g_{target}: \mathcal{X}_j \rightarrow [0,1],$ for all $j$.
        In this study, we use a self-masking MNAR mechanism, i.e., the missingness of a variable depends on its value alone. More precisely, we use an upper quantile censorship approach. The quantile level $q$ is chosen such that when missing values are generated on the $q$-quantile at random, the requested proportion of missing values is achieved. For more details about this chosen approach, we refer to the documentation of the \texttt{produce\_NA} function.\footnote{\url{https://rmisstastic.netlify.app/how-to/generate/missSimul.pdf}}
\end{enumerate}

\subsection{Summary of the data generation under the two alternative scenarios}

\begin{minipage}{0.48\textwidth}
\begin{algorithm}[H]
\label{algo:simulation-design-standard}
\SetAlgoLined
\KwResult{Joint data table $X^*$ of RCT and observational data and additional variables $A,Y$ for the RCT.}
Sample $N>>n$ observations $X_1, \dots, X_N$ from the target population $\mathcal{P}_{target}(X)$\; 
Sample $S$ according to the selection model (\eqref{eq:selection-model} or \eqref{eq:selection-model-nonlin}) on $X$\;
Keep only the $\{S=1\}$ indexed observations $X_{\mathcal{R}} \gets X_{\{i:\,S_i=1\}}$ as the RCT\; 
Sample $A$ according to a Bernoulli distribution $\mathcal{B}(0.5)$ (coin flip)\;
Sample $Y$ according to the outcome model (\eqref{eq:outcome-model} or \eqref{eq:outcome-model-nonlin})  on $X_{\mathcal{R}}$\;
Sample $m$ observations $X_{\mathcal{O}}$ from the target population $\mathcal{P}_{target}(X)$ as the observational data\; 
Concatenate the datasets: $X \gets [X_{\mathcal{R}}^T \,,\; X_{\mathcal{O}}^T]^T$ and append the indicator $\mathds{1}_{Set_{\mathcal{R}}}$ to the data ($X \gets [X\,,\; \mathds{1}_{Set_{\mathcal{R}}}]$)\; 
Sample missing values for the $n+m$ observations according to either \eqref{eq:mcar-simulation-mi}, \eqref{eq:mar-simulation-mi} or \eqref{eq:mnar-simulation-mi}\;
 \caption{Steps for simulation design under the standard assumption}
\end{algorithm}
\vspace{50pt}
\end{minipage}
\hfill
\begin{minipage}{0.48\textwidth}
\begin{algorithm}[H]
\label{algo:simulation-design-cis}
\SetAlgoLined
\KwResult{Joint data table $X^*$ of RCT and observational data and additional variables $A,Y$ for the RCT.}
Sample $N>>n$ observations $X_1, \dots, X_N$ from the target population $\mathcal{P}_{target}(X)$\;
Sample missing values for the $N$ observations according to either \eqref{eq:mcar-simulation-mi}, \eqref{eq:mar-simulation-mi} or \eqref{eq:mnar-simulation-mi}\;
Sample $S$ according to pattern-dependent selection model (\eqref{eq:selection-model-cis} or \eqref{eq:selection-model-cis-nonlin}) on $X$\;
Keep only the $\{S=1\}$ indexed observations $X_{\mathcal{R}} \gets X_{\{i:\,S_i=1\}}$ as the RCT\; 
Sample $A$ according to a Bernoulli distribution $\mathcal{B}(0.5)$ (coin flip)\;
Sample $Y$ according to the outcome model (\eqref{eq:outcome-model} or \eqref{eq:outcome-model-nonlin}) on $X_{\mathcal{R}}$\;
Sample $m$ observations $X_{\mathcal{O}}$ from the target population $\mathcal{P}_{target}(X)$ as the observational data\;
Sample missing values for the $m$ observations $X_{\mathcal{O}}$ using the same mechanism as before but possibly with different proportions\;
Concatenate $X_{\mathcal{R}}^*$ and $X_{\mathcal{O}}^*$ ($X^* \gets [X_{\mathcal{R}}^{*,T} \,,\; X_{\mathcal{O}}^{*,T}]^T$), and append the indicator $\mathds{1}_{Set_{\mathcal{R}}}$ to the data ($X^* \gets [X^*\,,\; \mathds{1}_{Set_{\mathcal{R}}}]$)\;
 \caption{Steps for simulation design under the CIS assumption.}
\end{algorithm}
\end{minipage}

\subsection{Results for the non-linear simulation setting.}

\begin{figure}[H]
\centering
   \includegraphics[width=0.95\textwidth]{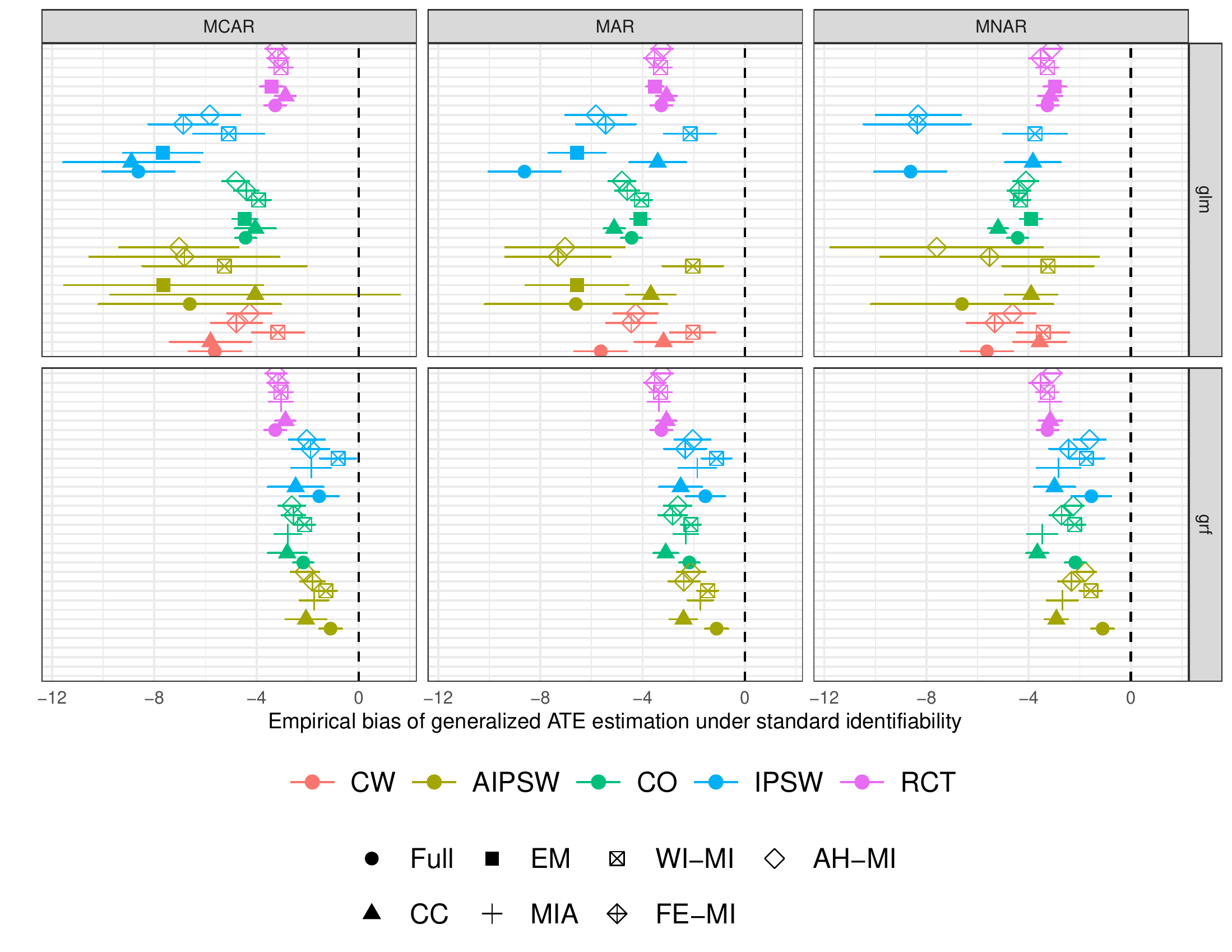}
\caption{Empirical bias of generalizing ATE estimators in the non-linear setting under the \protect\textit{standard CATE transportability assumption} \protect\ref{a:trans-2}, 95\% Monte Carlo confidence intervals, $n=2000$. Different colors denote different generalization estimators; different shapes represent different strategies to handle missing values. Full: estimation on full data; CC: estimation on complete cases; EM: estimation by expectation maximization directly handling missing values; MIA: estimation by generalized random forests with missing incorporated in attributes; WI-MI: within-study multiple imputation; FE-MI: fixed effect joint multiple imputation; AH-MI: ad-hoc joint multiple imputation.}
\label{fig:standard-bias-nsim100-nonlin}
\end{figure}

\begin{figure}[H]
    \centering
	\includegraphics[width=0.95\textwidth]{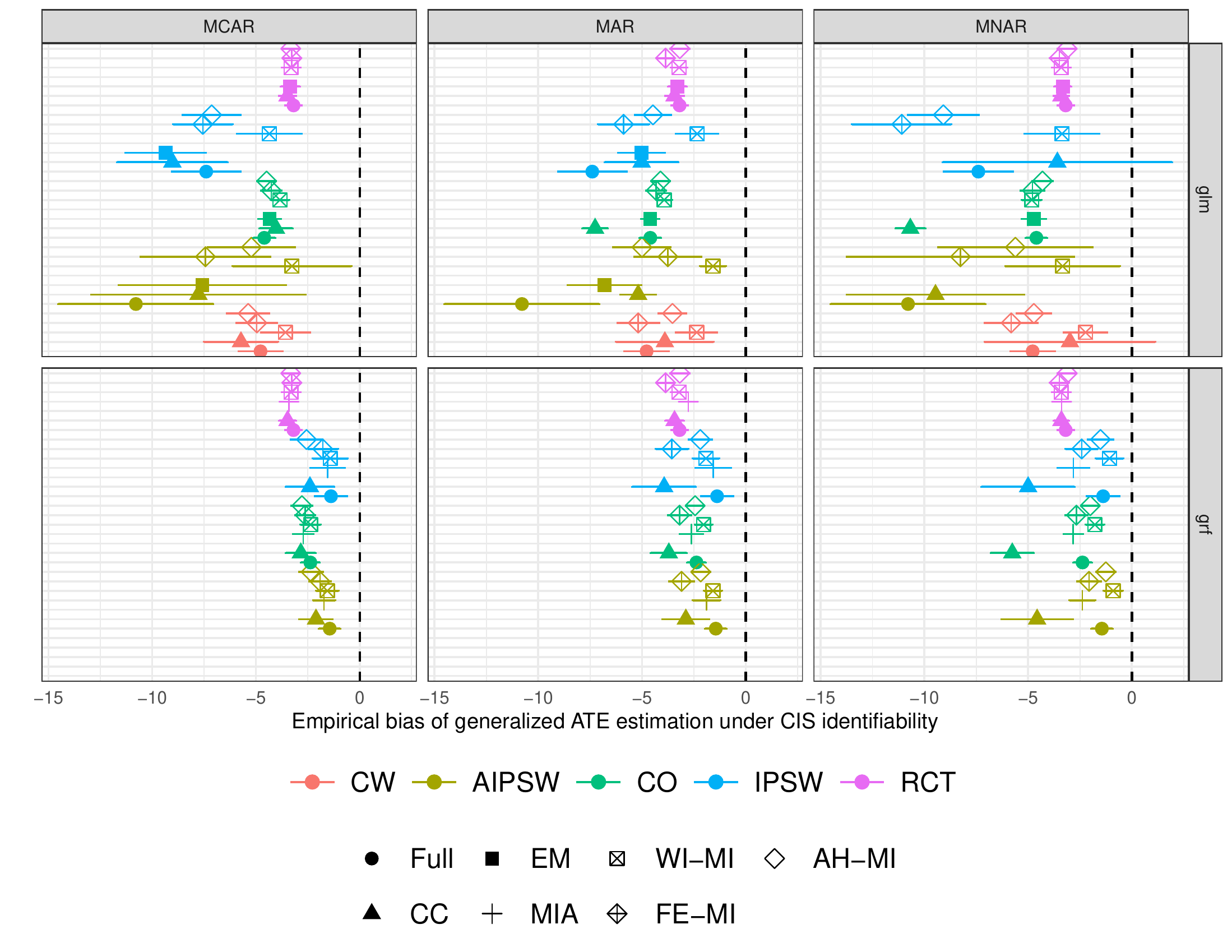}
\caption{Empirical bias of generalizing ATE estimators in the non-linear setting under the \protect\textit{alternative CATE transportability assumption (CIS)} \protect\ref{a:trans-2-na}, 95\% Monte Carlo confidence intervals, $n=2000$. Different colors denote different generalization estimators; different shapes represent different strategies to handle missing values. Full: estimation on full data; CC: estimation on complete cases; EM: estimation by expectation maximization directly handling missing values; MIA: estimation by generalized random forests with missing incorporated in attributes; WI-MI: within-study multiple imputation; FE-MI: fixed effect joint multiple imputation; AH-MI: ad-hoc joint multiple imputation.}
\label{fig:cis-bias-nsim100-nonlin}
\end{figure}

In this section, we report the detailed simulation results in the non-linear setting in Figures~\ref{fig:standard-bias-nsim100-nonlin} and \ref{fig:cis-bias-nsim100-nonlin}.

We observe in the first row of Figures~\ref{fig:standard-bias-nsim100-nonlin} and \ref{fig:cis-bias-nsim100-nonlin}, that under both categories of identifiability assumptions, all parametric estimators (based on logistic and linear regressions) fail to recover the true ATE $\tau$ and the bias of the estimators of the generalized effect even exceeds the ``true'' bias of the RCT sample ATE estimator (the difference in mean estimator for the RCT sample and therefore estimating $\tau_{trial}$).
Surprisingly however, this is also true for the non-parametric estimators (based on random forest regression models) shown in the second row of the same figures. While the bias of the non-parametric estimators is smaller than for the parametric estimators, as expected from random forests that are known for their good performance in the presence of non-linearities and high order interaction terms \citep{breiman2001random}, it is non-zero for all estimators, even the methods relying on the full data. An important observation is that the multiple imputation and the MIA-based estimators perform almost or at least as well as the full data estimator, with the within-study multiple imputation (WI-MI) performing best throughout the different simulation scenarios.
A plausible explanation for the poor performance of all estimators in the non-linear setting is the sample size which is moderately small for the chosen regression method with slow convergence rate. Indeed, the reported results refer to a sample size of $n=2000$ for the RCT and $m=20000$ for the observational sample. When choosing a sample size 5 times larger, then the full data AIPSW estimator is indeed unbiased as expected (results not reported here). Furthermore, the studied estimators achieve similar bias under the adequate identifiability assumptions (Assumptions \ref{a:trans-2} and \ref{a:pos} for the multiple imputation based estimators and Assumptions \ref{a:trans-2-na} and \ref{a:pos-na} for the MIA based estimators).

\section{Details on the critical care management application}\label{appendix:analysis}

\paragraph{Selection diagram of the CRASH-2 trial}

\begin{figure}[ht!]
\centering
\includegraphics[width=.7\textwidth]{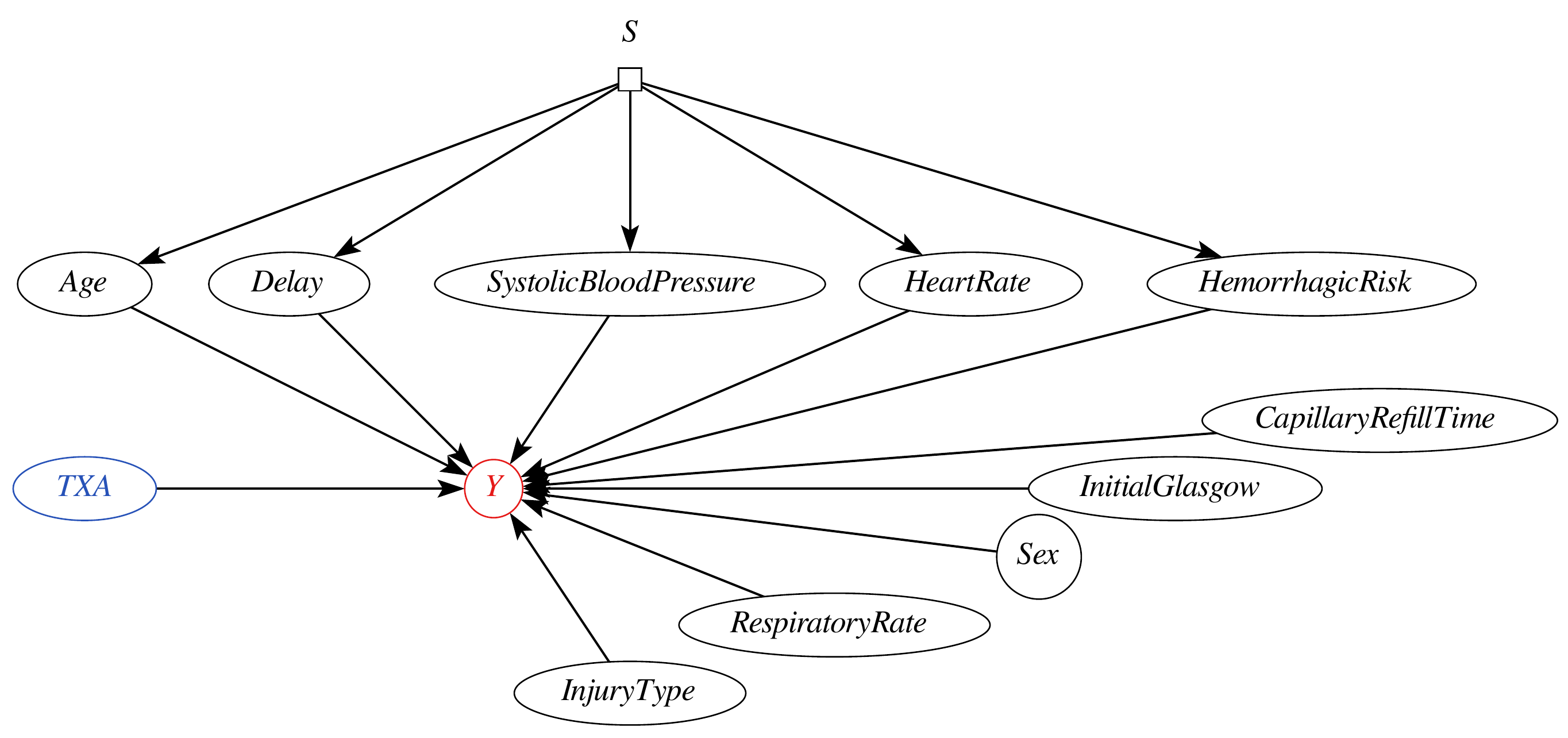}
\caption{Causal graph of CRASH-2 trial representing treatment, outcome, inclusion criteria with $S$ and other predictors of outcome (Figure generated using the Causal Fusion software by \citet{Bareinboim2016}).}
\label{fig:crash2-txa-dag}
\end{figure}

In Figure~\ref{fig:crash2-txa-dag}, we represent the selection variable $S$, the treatment assignment $TXA$ and the outcome $Y$, as well as the related covariates from the CRASH-2 trial in form of a causal diagram.

\begin{figure}[ht!]
    \centering
    \includegraphics[width=0.9\textwidth]{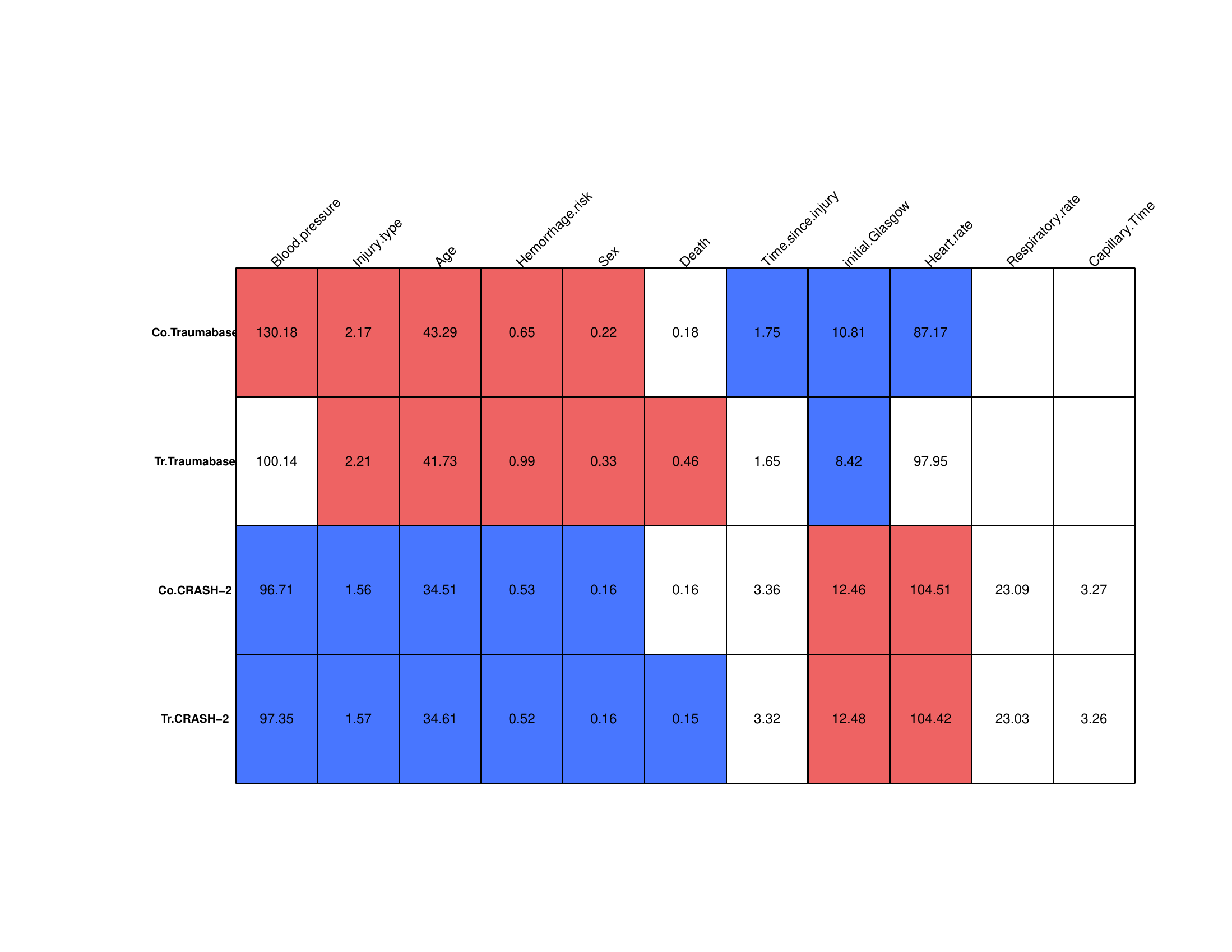}
    \caption{Distributional shift and difference in terms of univariate means of the trial inclusion criteria (red: group mean greater than overall mean that is taken over both data sets, blue: group mean less than overall mean, white: no significant difference with overall mean, numeric values: group mean (resp. proportion for categorical variables). For categorical variables a chi-squared test is used, for continuous variables a Fisher test (in a one-way analysis of variance). ``Tr": treatment group, ``Co": control group. Figure obtained with the \texttt{catdes} function of the \texttt{FactoMineR} package \citep{FactoMineR}.} 
    \label{fig:catdes_eligibility}
\end{figure}

\paragraph{Distributional shift}
The graph in Figure~\ref{fig:catdes_eligibility} illustrates the fundamental difference between the two studies, namely the treatment bias in the observational study and the balanced treatment groups in the RCT, but also a covariate shift between the two studies. 
 
\begin{figure}[ht!]
    \centering
    \includegraphics[width = \textwidth]{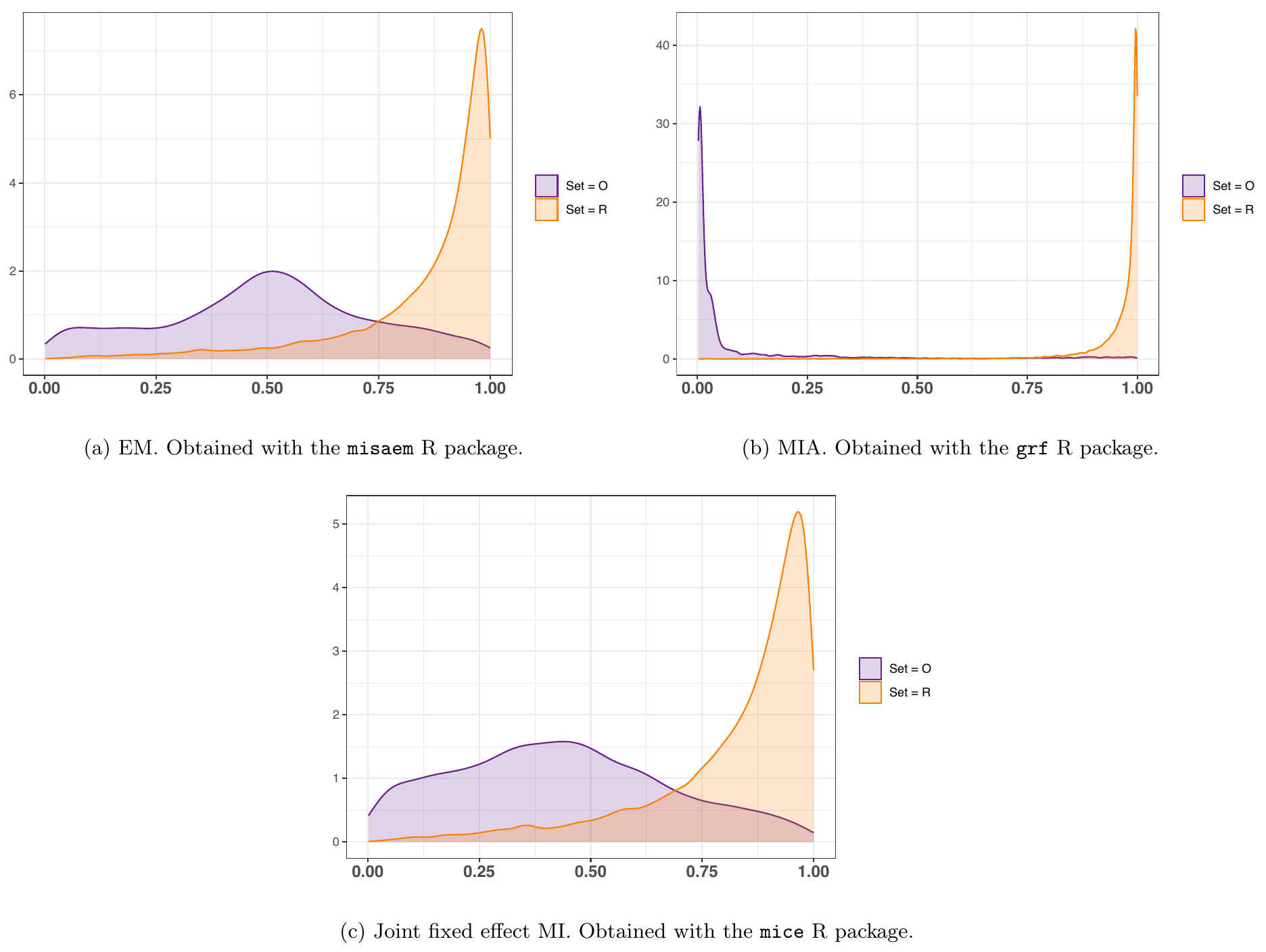}
    \caption{Estimated densities of the fitted selection scores.}
    \label{fig:propensityscores}
\end{figure}

\paragraph{Selection scores}
The distribution of the estimated selection scores are given in Figures~\ref{fig:propensityscores} (a), \ref{fig:propensityscores} (b), and \ref{fig:propensityscores} (c). We notice that the scores obtained using EM and MI+logistic regression are similar and suggest that the positivity assumption is satisfied since we observe a good degree of overlap between the distributions of the scores for the two data sets. The scores estimated via MIA however concentrate around 0 and 1 for the observational and RCT observations respectively, suggesting poor overlap under this model. In Appendix~\ref{appendix:analysis}, we provide further comparisons of the estimated selection scores, pointing towards the multiple imputation strategy as the suited approach in this case. 

These results provide an additional argument in favor of the multiple imputation strategy which appears to be more adapted to the handling of missing values in this analysis; in particular we will apply a joint fixed effect multiple imputation strategy since it outperforms the other multiple imputation strategies in the simulation study of Section~\ref{sec:simulations}.

\begin{figure}[ht!]
    \centering
    \includegraphics[width =\textwidth]{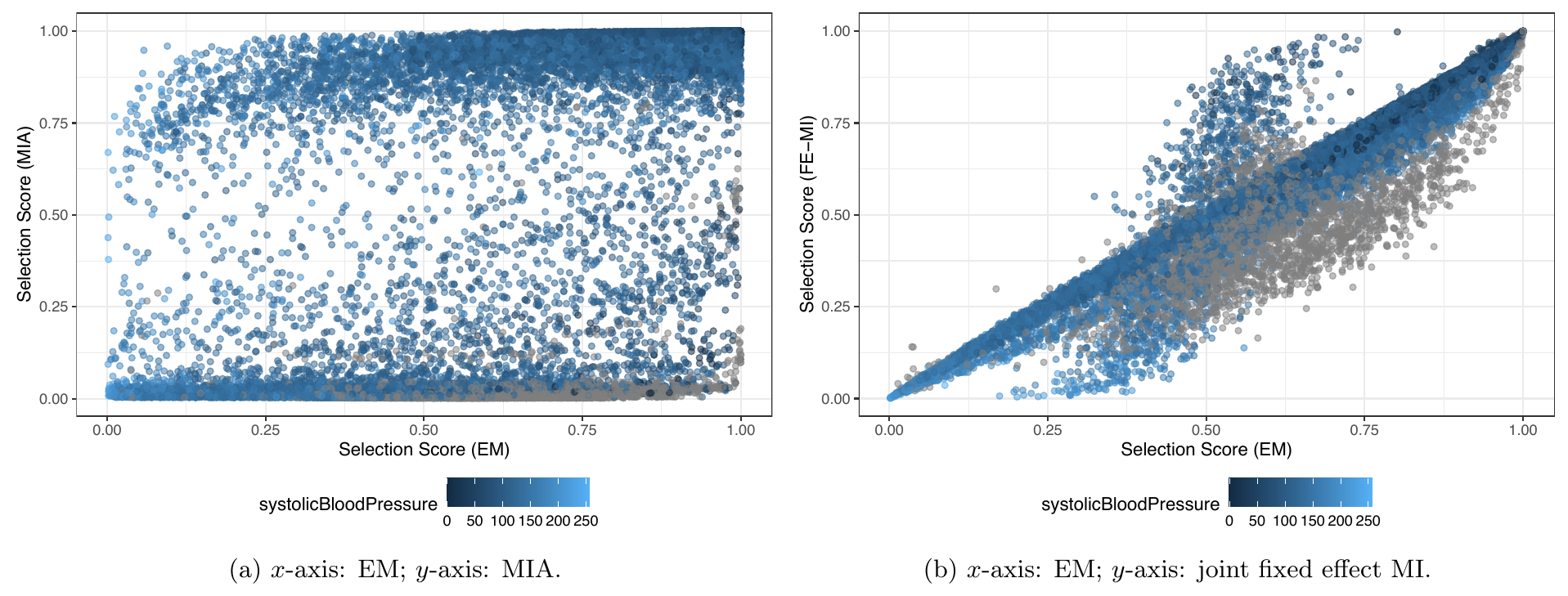}
    \caption{Scatter plots of different estimated selection scores. The point color is set according to the systolic blood pressure (SBP) covariate values (missing SBP values are indicated by gray points).}
       \label{fig:scatter_plot_propensity_score}
\end{figure}

We have seen in Section~\ref{sec:data-analysis} that the scores obtained using EM and MI+logistic regression suggest that the positivity assumption is satisfied, while the scores estimated via MIA however concentrate around 0 and 1 for the observational and RCT observations respectively, suggesting poor overlap under this model.
This can be also observed in the scatter plots in Figure~\ref{fig:scatter_plot_propensity_score}, where we color the observations according to the values of the systolic blood pressure (SBP, a variable with an important fraction of missing values in the observational data, see Figure~\ref{fig:percentagenaeligibility}). We notice that MIA attributes a very low selection score to all observations with missing SBP value and thus selects the response indicator of the SBP variable to partly predict trial eligibility. This method has been studied more extensively in a regression framework and not in a classification framework and here we find that it predicts a class according to the presence of missing values and this is not necessarily what we intend when applying this method.

\bibliographystyle{apalike}
\bibliography{biometrical_2023.bib}

\begin{thebibliography}{}

\bibitem[\protect\citeauthoryear{Athey, Tibshirani, and Wager}{Athey
  et~al.}{2019}]{athey_etal_AS2019}
Athey, S., J.~Tibshirani, and S.~Wager (2019).
\newblock Generalized random forests.
\newblock {\em The Annals of Statistics\/}~{\em 47\/}(2), 1148--1178.

\bibitem[\protect\citeauthoryear{Bareinboim and Pearl}{Bareinboim and
  Pearl}{2016}]{Bareinboim2016}
Bareinboim, E. and J.~Pearl (2016).
\newblock Causal inference and the data-fusion problem.
\newblock {\em Proceedings of the National Academy of Sciences\/}~{\em 113},
  7345--7352.

\bibitem[\protect\citeauthoryear{Bartlett, Harel, and Carpenter}{Bartlett
  et~al.}{2015}]{bartlett_etal_2015}
Bartlett, J.~W., O.~Harel, and J.~R. Carpenter (2015).
\newblock Asymptotically unbiased estimation of exposure odds ratios in
  complete records logistic regression.
\newblock {\em American journal of epidemiology\/}~{\em 182\/}(8), 730--736.

\bibitem[\protect\citeauthoryear{Blake, Leyrat, Mansfield, Tomlinson,
  Carpenter, and Williamson}{Blake et~al.}{2020}]{blake_etal_2019}
Blake, H.~A., C.~Leyrat, K.~E. Mansfield, L.~A. Tomlinson, J.~Carpenter, and
  E.~J. Williamson (2020).
\newblock Estimating treatment effects with partially observed covariates using
  outcome regression with missing indicators.
\newblock {\em Biometrical Journal\/}~{\em 62\/}(2), 428--443.

\bibitem[\protect\citeauthoryear{Breiman}{Breiman}{2001}]{breiman2001random}
Breiman, L. (2001).
\newblock Random forests.
\newblock {\em Machine learning\/}~{\em 45\/}(1), 5--32.

\bibitem[\protect\citeauthoryear{Cap}{Cap}{2019}]{crash3}
Cap, A.~P. (2019).
\newblock Crash-3: a win for patients with traumatic brain injury.
\newblock {\em The Lancet\/}~{\em 394\/}(10210), 1687 -- 1688.

\bibitem[\protect\citeauthoryear{Chernozhukov, Chetverikov, Demirer, Duflo,
  Hansen, Newey, and Robins}{Chernozhukov
  et~al.}{2018}]{chernozhukov_etal_2018}
Chernozhukov, V., D.~Chetverikov, M.~Demirer, E.~Duflo, C.~Hansen, W.~Newey,
  and J.~Robins (2018).
\newblock Double/debiased machine learning for treatment and structural
  parameters.
\newblock {\em The Econometrics Journal\/}~{\em 21\/}(1), C1--C68.

\bibitem[\protect\citeauthoryear{Chu, Lu, and Yang}{Chu
  et~al.}{2022}]{chu2022targeted}
Chu, J., W.~Lu, and S.~Yang (2022).
\newblock Targeted optimal treatment regime learning using summary statistics.
\newblock {\em arXiv preprint arXiv:2201.06229\/}.

\bibitem[\protect\citeauthoryear{Cole and Stuart}{Cole and
  Stuart}{2010}]{Cole2010}
Cole, S.~R. and E.~A. Stuart (2010).
\newblock Generalizing evidence from randomized clinical trials to target
  populations: The {ACTG} 320 trial.
\newblock {\em American Journal of Epidemiology\/}~{\em 172}, 107--115.

\bibitem[\protect\citeauthoryear{Colnet, Josse, Varoquaux, and Scornet}{Colnet
  et~al.}{2022}]{colnet_etal_2021}
Colnet, B., J.~Josse, G.~Varoquaux, and E.~Scornet (2022).
\newblock Causal effect on a target population: A sensitivity analysis to
  handle missing covariates.
\newblock {\em Journal of Causal Inference\/}~{\em 10\/}(1), 372--414.

\bibitem[\protect\citeauthoryear{Colnet, Mayer, Chen, Dieng, Li, Varoquaux,
  Vert, Josse, and Yang}{Colnet et~al.}{2020}]{Colnet2020}
Colnet, B., I.~Mayer, G.~Chen, A.~Dieng, R.~Li, G.~Varoquaux, J.-P. Vert,
  J.~Josse, and S.~Yang (2020).
\newblock Causal inference methods for combining randomized trials and
  observational studies: a review.
\newblock {\em arXiv preprint arXiv:2011.08047\/}.

\bibitem[\protect\citeauthoryear{{{CRASH-2} Collaborators} et~al.}{{{CRASH-2}
  Collaborators} et~al.}{2011}]{Crash2011}
{{CRASH-2} Collaborators} et~al. (2011).
\newblock The importance of early treatment with tranexamic acid in bleeding
  trauma patients: an exploratory analysis of the crash-2 randomised controlled
  trial.
\newblock {\em The Lancet\/}~{\em 377\/}(9771), 1096--1101.

\bibitem[\protect\citeauthoryear{D'Agostino and Rubin}{D'Agostino and
  Rubin}{2000}]{dagostino_rubin_JASA2000}
D'Agostino, Jr, R.~B. and D.~B. Rubin (2000).
\newblock Estimating and using propensity scores with partially missing data.
\newblock {\em Journal of the American Statistical Association\/}~{\em
  95\/}(451), 749--759.

\bibitem[\protect\citeauthoryear{Dahabreh and Hern{\'a}n}{Dahabreh and
  Hern{\'a}n}{2019}]{Dahabreh2019}
Dahabreh, I.~J. and M.~A. Hern{\'a}n (2019).
\newblock Extending inferences from a randomized trial to a target population.
\newblock {\em European Journal of Epidemiology\/}~{\em 34\/}(8), 719--722.

\bibitem[\protect\citeauthoryear{Dahabreh, Robertson, and Hern{\'a}n}{Dahabreh
  et~al.}{2019}]{Dahabreh2019c}
Dahabreh, I.~J., S.~E. Robertson, and M.~A. Hern{\'a}n (2019).
\newblock On the relation between g-formula and inverse probability weighting
  estimators for generalizing trial results.
\newblock {\em Epidemiology\/}~{\em 30\/}(6), 807--812.

\bibitem[\protect\citeauthoryear{Degtiar and Rose}{Degtiar and
  Rose}{2023}]{degtiar2021review}
Degtiar, I. and S.~Rose (2023).
\newblock A review of generalizability and transportability.
\newblock {\em Annual Review of Statistics and Its Application\/}~{\em
  10\/}(1).

\bibitem[\protect\citeauthoryear{Dempster, Laird, and Rubin}{Dempster
  et~al.}{1977}]{dempster_etal_1977}
Dempster, A.~P., N.~M. Laird, and D.~B. Rubin (1977).
\newblock Maximum likelihood from incomplete data via the em algorithm.
\newblock {\em Journal of the Royal Statistical Society: Series B
  (Methodological)\/}~{\em 39\/}(1), 1--22.

\bibitem[\protect\citeauthoryear{Efron and Tibshirani}{Efron and
  Tibshirani}{1994}]{Efron1994}
Efron, B. and R.~J. Tibshirani (1994).
\newblock {\em An introduction to the bootstrap}.
\newblock CRC press.

\bibitem[\protect\citeauthoryear{Hamada, Rosa, Gauss, Desclefs, Raux, Harrois,
  Follin, Cook, Boutonnet, Attias, et~al.}{Hamada
  et~al.}{2018}]{hamada2018development}
Hamada, S.~R., A.~Rosa, T.~Gauss, J.-P. Desclefs, M.~Raux, A.~Harrois,
  A.~Follin, F.~Cook, M.~Boutonnet, A.~Attias, et~al. (2018).
\newblock Development and validation of a pre-hospital “red flag” alert for
  activation of intra-hospital haemorrhage control response in blunt trauma.
\newblock {\em Critical Care\/}~{\em 22\/}(1), 1--12.

\bibitem[\protect\citeauthoryear{Hern{\'a}n and Robins}{Hern{\'a}n and
  Robins}{2016}]{hernan2016using}
Hern{\'a}n, M.~A. and J.~M. Robins (2016).
\newblock Using big data to emulate a target trial when a randomized trial is
  not available.
\newblock {\em American journal of epidemiology\/}~{\em 183\/}(8), 758--764.

\bibitem[\protect\citeauthoryear{Hijazi, Abu~Fanne, Abramovitch, Yarovoi,
  Higazi, Abdeen, Basheer, Maraga, Cines, and Al-Roof~Higazi}{Hijazi
  et~al.}{2015}]{Hijazi2015}
Hijazi, N., R.~Abu~Fanne, R.~Abramovitch, S.~Yarovoi, M.~Higazi, S.~Abdeen,
  M.~Basheer, E.~Maraga, D.~B. Cines, and A.~Al-Roof~Higazi (2015).
\newblock Endogenous plasminogen activators mediate progressive intracerebral
  hemorrhage after traumatic brain injury in mice.
\newblock {\em Blood, The Journal of the American Society of Hematology\/}~{\em
  125\/}(16), 2558--2567.

\bibitem[\protect\citeauthoryear{Hippel}{Hippel}{2009}]{Hippel2009}
Hippel, P.~v. (2009).
\newblock How to impute interactions, squares, and other transformed variables.
\newblock {\em Sociological Methodology\/}~{\em 39\/}(1), 265--291.

\bibitem[\protect\citeauthoryear{Imbens and Menzel}{Imbens and
  Menzel}{2021}]{imbens2018causal}
Imbens, G. and K.~Menzel (2021).
\newblock A causal bootstrap.
\newblock {\em The Annals of Statistics\/}~{\em 49\/}(3), 1460--1488.

\bibitem[\protect\citeauthoryear{Jiang, Josse, Lavielle, and Group}{Jiang
  et~al.}{2020}]{jiang_etal_2018}
Jiang, W., J.~Josse, M.~Lavielle, and T.~Group (2020).
\newblock Logistic regression with missing covariates—parameter estimation,
  model selection and prediction within a joint-modeling framework.
\newblock {\em Computational Statistics \& Data Analysis\/}~{\em 145}, 106907.

\bibitem[\protect\citeauthoryear{Josse, Prost, Scornet, and Varoquaux}{Josse
  et~al.}{2019}]{josse_etal_2019}
Josse, J., N.~Prost, E.~Scornet, and G.~Varoquaux (2019).
\newblock On the consistency of supervised learning with missing values.
\newblock {\em arXiv preprint arXiv:1902.06931\/}.

\bibitem[\protect\citeauthoryear{Josse and Reiter}{Josse and
  Reiter}{2018}]{josse2018}
Josse, J. and J.~P. Reiter (2018, 05).
\newblock Introduction to the special section on missing data.
\newblock {\em Statist. Sci.\/}~{\em 33\/}(2), 139--141.

\bibitem[\protect\citeauthoryear{Kallus, Mao, and Udell}{Kallus
  et~al.}{2018}]{kallus_etal_2018}
Kallus, N., X.~Mao, and M.~Udell (2018).
\newblock Causal inference with noisy and missing covariates via matrix
  factorization.
\newblock In {\em Advances in Neural Information Processing Systems}, pp.\
  6921--6932.

\bibitem[\protect\citeauthoryear{Kim and Shao}{Kim and Shao}{2013}]{Kim2013b}
Kim, J.~K. and J.~Shao (2013).
\newblock {\em Statistical methods for handling incomplete Data}.
\newblock CRC Press.

\bibitem[\protect\citeauthoryear{L\^e, Josse, and Husson}{L\^e
  et~al.}{2008}]{FactoMineR}
L\^e, S., J.~Josse, and F.~Husson (2008).
\newblock {FactoMineR}: A package for multivariate analysis.
\newblock {\em Journal of Statistical Software\/}~{\em 25\/}(1), 1--18.

\bibitem[\protect\citeauthoryear{Le~Morvan, Josse, Moreau, Scornet, and
  Varoquaux}{Le~Morvan et~al.}{2020}]{LeMorvan2020}
Le~Morvan, M., J.~Josse, T.~Moreau, E.~Scornet, and G.~Varoquaux (2020).
\newblock Neumiss networks: differential programming for supervised learning
  with missing values.
\newblock In {\em Advances in Neural Information Processing Systems 33}.

\bibitem[\protect\citeauthoryear{Le~Morvan, Josse, Scornet, and
  Varoquaux}{Le~Morvan et~al.}{2021}]{LeMorvan2021}
Le~Morvan, M., J.~Josse, E.~Scornet, and G.~Varoquaux (2021).
\newblock What’sa good imputation to predict with missing values?
\newblock {\em Advances in Neural Information Processing Systems\/}~{\em 34}.

\bibitem[\protect\citeauthoryear{Lee, Yang, Dong, Wang, Zeng, and Cai}{Lee
  et~al.}{2022}]{dong2020integrative}
Lee, D., S.~Yang, L.~Dong, X.~Wang, D.~Zeng, and J.~Cai (2022).
\newblock Improving trial generalizability using observational studies.
\newblock {\em Biometrics\/}, 1--13.

\bibitem[\protect\citeauthoryear{Leigh, Collaborators, Guo, Deribe, Brazinova,
  and Hostiuc}{Leigh et~al.}{2018}]{who2017}
Leigh, J., G.~Collaborators, Y.~Guo, K.~Deribe, A.~Brazinova, and S.~Hostiuc
  (2018, 11).
\newblock Global,regional,and national disability- adjusted life years(dalys)
  for 359 diseases and injuries and healthy life expectancy(hale) for 195
  countries and territories,1990-2017: a systematic analysis for the global
  burden of disease study 2017.
\newblock {\em The Lancet\/}~{\em 392}, 1859--1922.

\bibitem[\protect\citeauthoryear{Leyrat, Seaman, White, Douglas, Smeeth, Kim,
  Resche-Rigon, Carpenter, and Williamson}{Leyrat
  et~al.}{2019}]{leyrat_etal_2019}
Leyrat, C., S.~R. Seaman, I.~R. White, I.~Douglas, L.~Smeeth, J.~Kim,
  M.~Resche-Rigon, J.~R. Carpenter, and E.~J. Williamson (2019).
\newblock Propensity score analysis with partially observed covariates: How
  should multiple imputation be used?
\newblock {\em Statistical methods in medical research\/}~{\em 28\/}(1), 3--19.

\bibitem[\protect\citeauthoryear{Li, Morgan, and Zaslavsky}{Li
  et~al.}{2018}]{li_etal_JASA2018}
Li, F., K.~L. Morgan, and A.~M. Zaslavsky (2018).
\newblock Balancing covariates via propensity score weighting.
\newblock {\em Journal of the American Statistical Association\/}~{\em
  113\/}(521), 390--400.

\bibitem[\protect\citeauthoryear{Little and Badawy}{Little and
  Badawy}{2019}]{little2019causal}
Little, M.~A. and R.~Badawy (2019).
\newblock Causal bootstrapping.
\newblock {\em arXiv preprint arXiv:1910.09648\/}.

\bibitem[\protect\citeauthoryear{Little and Rubin}{Little and
  Rubin}{2014}]{Little2014}
Little, R.~J. and D.~B. Rubin (2014).
\newblock {\em {Statistical Analysis with Missing Data}}.
\newblock New York: Wiley.

\bibitem[\protect\citeauthoryear{Mattei}{Mattei}{2009}]{mattei_mealli_SMA2009}
Mattei, A. (2009).
\newblock Estimating and using propensity score in presence of missing
  background data: an application to assess the impact of childbearing on
  wellbeing.
\newblock {\em Statistical Methods and Applications\/}~{\em 18\/}(2), 257--273.

\bibitem[\protect\citeauthoryear{Mayer, Josse, Tierney, and Vialaneix}{Mayer
  et~al.}{2022}]{mayer_etal_2019}
Mayer, I., J.~Josse, N.~Tierney, and N.~Vialaneix (2022).
\newblock R-miss-tastic: a unified platform for missing values methods and
  workflows.
\newblock {\em R Journal\/}~{\em 14\/}(2), 244--266.

\bibitem[\protect\citeauthoryear{Mayer, Sverdrup, Gauss, Moyer, Wager, and
  Josse}{Mayer et~al.}{2020}]{Mayer2020}
Mayer, I., E.~Sverdrup, T.~Gauss, J.-D. Moyer, S.~Wager, and J.~Josse (2020).
\newblock Doubly robust treatment effect estimation with missing attributes.
\newblock {\em Ann. Appl. Statist.\/}~{\em 14\/}(3), 1409--1431.

\bibitem[\protect\citeauthoryear{Mealli and Rubin}{Mealli and
  Rubin}{2015}]{mealli2015clarifying}
Mealli, F. and D.~B. Rubin (2015).
\newblock Clarifying missing at random and related definitions, and
  implications when coupled with exchangeability.
\newblock {\em Biometrika\/}~{\em 102\/}(4), 995--1000.

\bibitem[\protect\citeauthoryear{Morris, White, and Crowther}{Morris
  et~al.}{2019}]{morris2019using}
Morris, T.~P., I.~R. White, and M.~J. Crowther (2019).
\newblock Using simulation studies to evaluate statistical methods.
\newblock {\em Statistics in medicine\/}~{\em 38\/}(11), 2074--2102.

\bibitem[\protect\citeauthoryear{Nie, Imbens, and Wager}{Nie
  et~al.}{2021}]{nie2021covariate}
Nie, X., G.~Imbens, and S.~Wager (2021).
\newblock Covariate balancing sensitivity analysis for extrapolating randomized
  trials across locations.
\newblock {\em arXiv preprint arXiv:2112.04723\/}.

\bibitem[\protect\citeauthoryear{Robins}{Robins}{1986}]{Robins1986}
Robins, J. (1986).
\newblock A new approach to causal inference in mortality studies with a
  sustained exposure period--application to control of the healthy worker
  survivor effect.
\newblock {\em Mathematical Modelling\/}~{\em 7}, 1393--1512.

\bibitem[\protect\citeauthoryear{Roozenbeek, Maas, and Menon}{Roozenbeek
  et~al.}{2013}]{roozenbeek2013changing}
Roozenbeek, B., A.~I. Maas, and D.~K. Menon (2013).
\newblock Changing patterns in the epidemiology of traumatic brain injury.
\newblock {\em Nature Reviews Neurology\/}~{\em 9\/}(4), 231--236.

\bibitem[\protect\citeauthoryear{Rosenbaum and Rubin}{Rosenbaum and
  Rubin}{1984}]{rosenbaum_rubin_JASA1984}
Rosenbaum, P.~R. and D.~B. Rubin (1984).
\newblock Reducing bias in observational studies using subclassification on the
  propensity score.
\newblock {\em Journal of the American Statistical Association\/}~{\em
  79\/}(387), 516--524.

\bibitem[\protect\citeauthoryear{Rubin}{Rubin}{1976}]{rubin_1976}
Rubin, D.~B. (1976).
\newblock Inference and missing data.
\newblock {\em Biometrika\/}~{\em 63\/}(3), 581--592.

\bibitem[\protect\citeauthoryear{Rubin}{Rubin}{1987}]{Rubin1987}
Rubin, D.~B. (1987).
\newblock {\em Multiple Imputation for Nonresponse in Surveys}.
\newblock New York: Wiley.

\bibitem[\protect\citeauthoryear{Schafer}{Schafer}{2010}]{Schafer2010}
Schafer, J.~L. (2010).
\newblock {\em Analysis of incomplete multivariate data}.
\newblock London: Chapman and Hall: CRC press.

\bibitem[\protect\citeauthoryear{Seaman and White}{Seaman and
  White}{2014}]{seaman_white_2014}
Seaman, S. and I.~White (2014).
\newblock Inverse probability weighting with missing predictors of treatment
  assignment or missingness.
\newblock {\em Communications in Statistics-Theory and Methods\/}~{\em
  43\/}(16), 3499--3515.

\bibitem[\protect\citeauthoryear{Shakur-Still, Roberts, Bautista, Caballero,
  Coats, Dewan, El-Sayed, Tamar, Gupta, Herrera, Hunt, Iribhogbe, Izurieta,
  Khamis, Komolafe, Marrero, Mejía-Mantilla, Miranda, Uribe, and
  Yutthakasemsunt}{Shakur-Still et~al.}{2009}]{crash2}
Shakur-Still, H., I.~Roberts, R.~Bautista, J.~Caballero, T.~Coats, Y.~Dewan,
  H.~El-Sayed, G.~Tamar, S.~Gupta, J.~Herrera, B.~Hunt, P.~Iribhogbe,
  M.~Izurieta, H.~Khamis, E.~Komolafe, M.~Marrero, J.~Mejía-Mantilla, J.~J.
  Miranda, C.~Uribe, and S.~Yutthakasemsunt (2009, 11).
\newblock Effects of tranexamic acid on death, vascular occlusive events, and
  blood transfusion in trauma patients with significant haemorrhage
  ({CRASH-2}): A randomised, placebo-controlled trial.
\newblock {\em Lancet\/}~{\em 376}, 23--32.

\bibitem[\protect\citeauthoryear{Stuart, Cole, Bradshaw, and Leaf}{Stuart
  et~al.}{2011}]{Stuart2011}
Stuart, E.~A., S.~R. Cole, C.~P. Bradshaw, and P.~J. Leaf (2011).
\newblock The use of propensity scores to assess the generalizability of
  results from randomized trials.
\newblock {\em J. R. Statist. Soc. A\/}~{\em 174}, 369--386.

\bibitem[\protect\citeauthoryear{Tibshirani, Athey, Friedberg, Hadad,
  Hirshberg, Miner, Sverdrup, Wager, and Wright}{Tibshirani et~al.}{2020}]{grf}
Tibshirani, J., S.~Athey, R.~Friedberg, V.~Hadad, D.~Hirshberg, L.~Miner,
  E.~Sverdrup, S.~Wager, and M.~Wright (2020).
\newblock {\em grf: Generalized Random Forests}.
\newblock R package version 1.1.0.

\bibitem[\protect\citeauthoryear{Twala, Jones, and Hand}{Twala
  et~al.}{2008}]{twala_etal_2008}
Twala, B., M.~Jones, and D.~J. Hand (2008).
\newblock Good methods for coping with missing data in decision trees.
\newblock {\em Pattern Recognition Letters\/}~{\em 29\/}(7), 950--956.

\bibitem[\protect\citeauthoryear{{van Buuren}}{{van
  Buuren}}{2018}]{vanbuuren_2018}
{van Buuren}, S. (2018).
\newblock {\em Flexible Imputation of Missing Data. Second Edition}.
\newblock Boca Raton, FL: Chapman and Hall/CRC.

\bibitem[\protect\citeauthoryear{{van Buuren} and Groothuis-Oudshoorn}{{van
  Buuren} and Groothuis-Oudshoorn}{2011}]{mice}
{van Buuren}, S. and K.~Groothuis-Oudshoorn (2011).
\newblock {mice}: Multivariate imputation by chained equations in r.
\newblock {\em Journal of Statistical Software\/}~{\em 45\/}(3), 1--67.

\bibitem[\protect\citeauthoryear{Wager and Athey}{Wager and
  Athey}{2018}]{wager_athey_JASA2018}
Wager, S. and S.~Athey (2018).
\newblock Estimation and inference of heterogeneous treatment effects using
  random forests.
\newblock {\em Journal of the American Statistical Association\/}~{\em
  113\/}(523), 1228--1242.

\bibitem[\protect\citeauthoryear{Wutzler, Maegele, Marzi, Spanholtz, Wafaisade,
  Lefering, of~the German Society~for Trauma~Surgery, et~al.}{Wutzler
  et~al.}{2009}]{wutzler2009association}
Wutzler, S., M.~Maegele, I.~Marzi, T.~Spanholtz, A.~Wafaisade, R.~Lefering,
  T.~R. of~the German Society~for Trauma~Surgery, et~al. (2009).
\newblock Association of preexisting medical conditions with in-hospital
  mortality in multiple-trauma patients.
\newblock {\em Journal of the American College of Surgeons\/}~{\em 209\/}(1),
  75--81.

\bibitem[\protect\citeauthoryear{Yang, Wang, and Ding}{Yang
  et~al.}{2019}]{yang_etal_2019}
Yang, S., L.~Wang, and P.~Ding (2019).
\newblock Causal inference with confounders missing not at random.
\newblock {\em Biometrika\/}~{\em 106\/}(4), 875--888.

\bibitem[\protect\citeauthoryear{Zhu, Wang, and Samworth}{Zhu
  et~al.}{2022}]{zhu2019high}
Zhu, Z., T.~Wang, and R.~J. Samworth (2022).
\newblock High-dimensional principal component analysis with heterogeneous
  missingness.
\newblock {\em J. Roy. Statist. Soc., Ser. B\/}~{\em 84}, 2000--2031.

\end{thebibliography}

\end{document}